\documentclass[preprint,showpacs,preprintnumbers,amsmath,amssymb,nofootinbib]{revtex4}

\usepackage{amsthm}

{Corollary}[section]

\usepackage{etex}
\usepackage{amscd,amsbsy,array}
\usepackage{bm}
\usepackage{amsmath,dsfont}
\usepackage{soul} 
\usepackage{graphics,graphicx,xcolor}
\usepackage[colorlinks=true, pdfstartview=FitV, linkcolor=blue, citecolor=blue, urlcolor=blue]{hyperref} %
\usepackage{yfonts}

\newcommand{\red}{\textcolor{red}}
\newcommand{\blue}{\textcolor{blue}}

\newcommand{\orange}{\textcolor{orange}}
\newcommand{\gb}{\colorbox{green}}

\newcommand{\dgreen}{\textcolor[rgb]{0,0.35,0}}
\newcommand{\cyan}{\textcolor{cyan}}
\newcommand{\magenta}{\textcolor{magenta}}
\newcommand{\purple}{\textcolor{purple}}

\newenvironment{redtext}{\color{red}}{\ignorespacesafterend}
\newenvironment{bluetext}{\color{blue}}{\ignorespacesafterend}

\newenvironment{magentatext}{\color{magenta}}{\ignorespacesafterend}
\newenvironment{orangetext}{\color{orange}}{\ignorespacesafterend}
\newenvironment{cyantext}{\color{cyan}}{\ignorespacesafterend}
\newcommand{\bblue}{\begin{bluetext}}
\newcommand{\eblue}{\end{bluetext}}
\newcommand{\bred}{\begin{redtext}}
\newcommand{\ered}{\end{redtext}}
\newcommand{\bmagenta}{\begin{magentatext}}
\newcommand{\emagenta}{\end{magentatext}}
\newcommand{\borange}{\begin{orangetext}}
\newcommand{\eorange}{\end{orangetext}}
\newcommand{\bcyan}{\begin{cyantext}}
\newcommand{\ecyan}{\end{cyantext}}

\numberwithin{equation}{section}

\let\ssection=\section
\renewcommand{\section}{\setcounter{equation}{0}\ssection}

\def\bec{\begin{center}}
\def\ec{\end{center}}

\newcommand{\cA}{{\mathcal{A}}}

\newcommand{\GW}{{gravitational wave\;}}
\newcommand{\PT}{{P\"oschl{\strut}-Teller\;}}
\newcommand{\GWs}{{gravitational waves\;}}
\newcommand{\SL}{{Sturm-Liouville\;}}

\newcommand{\cI}{{\mathcal{I}}}

\newcommand{\cL}{{\mathcal{L}}}

\def\smallover#1/#2{\hbox{$\textstyle\frac{#1}{#2}$}} %

\def\parag{\hfil\break} 
\def\kikezd{\parag\underbar}

\def\besub{\begin{subequations}}
\def\esub{\end{subequations}}
\def\benu{\begin{enumerate}}
\def\eenu{\end{enumerate}}
\def\beq{\begin{equation}}
\def\eeq{\end{equation}}
\def\beqa{\begin{eqnarray}}
\def\eeqa{\end{eqnarray}}

\def\barray{\left(\begin{array}}
\def\earray{\end{array}\right)}
\def\barraynb{\begin{array}}
\def\earraynb{\end{array}}

\def\?{\quad{\gb{\fbox{\texttt{?}}\;}}\quad}
\def\p{{\partial}}

\def\v0{\mathbf{0}}

\def\Rarrow{{\quad\Rightarrow\quad}}

\def\beq{\begin{equation}}
\def\eeq{\end{equation}}
\def\bea{\begin{eqnarray}}
\def\eea{\end{eqnarray}}

\def\p{\partial}

\def \p{{\partial}}

\def\6{\partial}
\def\7{\tilde}
\def\8{\widehat}

\def\G11{\Gamma_{11} }

\newcommand{\const}{\mathop{\rm const.}\nolimits}
\newcommand{\half }{\frac{1}{2}}

\def\smallover#1/#2{\hbox{$\textstyle\frac{#1}{#2}$}} %
\def\smallcirc{{\raise 0.5pt \hbox{$\scriptstyle\circ$}}}
\def\2{{\smallover1/2}}
\def\aand{{\quad\text{\small and}\quad}}
\def\where{{\quad\text{\small where}\quad}}
\def\for{{\;\;\text{\small for}\;\;}}

\def\ie{{\;\text{\small i.e.}\;}}
\def\ie,{{\;\text{\small i.e.,}\;}}

\newcommand{\fm}{\mathfrak{m}}

\newcommand{\bigbox}[1]{\fbox{%
\rule[-20pt]{0pt}{45pt}$\;\;\displaystyle{#1}\;\;$}
}
\newcommand{\medbox}[1]{\fbox{%
\rule[-10pt]{0pt}{25pt}$\;\;\displaystyle{#1}\;\;$}%
}

\let\ssection=\section
\renewcommand{\section}{\setcounter{equation}{0}\ssection}

\begin{document}

\preprint{\texttt{arXiv:2405.12928v3 [gr-qc]}}

\title{Displacement within velocity effect\\
in \\
gravitational wave memory\footnote{Extended version of the talks given by PAH at the Wigner Institute in Budapest (April 5 2024), at the Workshop
{\sl Carrollian Physics and Holography}${}_{-}$CDFG${}_{-}$2024
at the Schr\"odinger Institute in Vienna (April 17  2024), and at the conference {\sl ``Conformal anomalies: theory and applications 2024,
https://indico.math.cnrs.fr/event/10718/} in Tours (May 7  2024).}}

\author{
P. M. Zhang$^{1}$\footnote{mailto:zhangpm5@mail.sysu.edu.cn},
P. A. Horvathy$^{2,3}$\footnote{mailto:horvathy@univ-tours.fr},
}

\affiliation{
$^1$ School of Physics and Astronomy, Sun Yat-sen University, Zhuhai, China
\\
${}^{2}$ Institut Denis-Poisson CNRS/UMR 7013 - Universit\'e de Tours - Universit\'e d'Orl\'eans Parc de Grammont, 37200; Tours, FRANCE 
\\
${}^{3}$ Erwin Schr\"odinger Institute, Vienna (Austria)
\\
\\}
\date{\today}

\pacs{
04.20.-q  Classical general relativity;\\
04.30.-w Gravitational waves
}

\begin{abstract}
Particles initially at rest hit by a passing sandwich gravitational wave exhibit, in general, the \emph{velocity  memory effect} (VM): they fly apart with constant velocity. For specific values of the wave parameters their motion can however become pure \emph{displacement} (DM) as suggested by Zel'dovich and Polnarev. For such a ``miraculous'' value, the particle trajectory is composed of an integer number of (approximate) standing half-waves.
Our statements are illustrated numerically by a Gaussian, and analytically by the P\"oschl-Teller profiles.
\bigskip

Annals of Physics \textbf{470} (2024) 169784\, https://doi.org.10.1016/j.aop.2024.169784
\end{abstract}

\maketitle

\tableofcontents

\section{Introduction}\label{Intro}

The  \emph{Memory Effect}, which may be a way to detect \GWs (GW), has two versions.
The \emph{displacement effect} (DM)
proposed by Zel'dovich and Polnarev \cite{ZelPol} for flyby suggests that  particles initially at rest are hit by a burst of \GW\!, \emph{although the distance between free bodies will change, their relative velocity will become vanishingly small as the flyby concludes}
  \cite{ZelPol,BraTho,BraGri}.

Confirmation could be obtained by taking into account non-linear effects \cite{Sachs61,Sachs62,Christo,BlaDam}.

Earlier studies \cite{Ehlers,Hawking68,GibbHawk71,Sou73,GriPol,AiBalasin,BoPi89}
 argued instead in favor of a \emph{velocity effect} ({VM}):  the particles would be scattered apart with \emph{constant velocity} by a burst of \GW\!.
Our previous investigations \cite{ShortMemory,LongMemory,PolPer,SLheart,EZHRev}
 confirmed {VM} however casted a doubt on the claim of Zel'dovich and Polnarev. In this paper we show that while {VM} is generic, \emph{pure displacement} \emph{may} indeed arise --- however only for exceptional values of the parameters ~: the trajectories must be composed of an \emph{integer number of half-waves} (reminiscent of quantum  conditions).
\goodbreak

We illustrate our statement by two closely related examples,
one numerical and the other analytical. Their profile is
 (i) either a Gaussian, \eqref{A0G}, or (ii) the P\"oschl-Teller  potential \cite{PTeller}, \eqref{PTPot}.
Their study sheds some light on the remarkable relation of the velocity and the displacement effects, {VM} and {DM}, respectively.
Our result is curiously related to zero-energy time-independent solutions of the Schr\"odinger equation.

In this paper we consider $D=1$ transverse dimension.
Our investigations will be extended to more physical  profiles and dimensions appropriate for flyby, gravitational collapse, etc in a follow-up paper \cite{DMvsVM}.
\goodbreak

\section{Memory effect}\label{MemorySec}

\begin{figure}[ht]
\includegraphics[scale=.45]{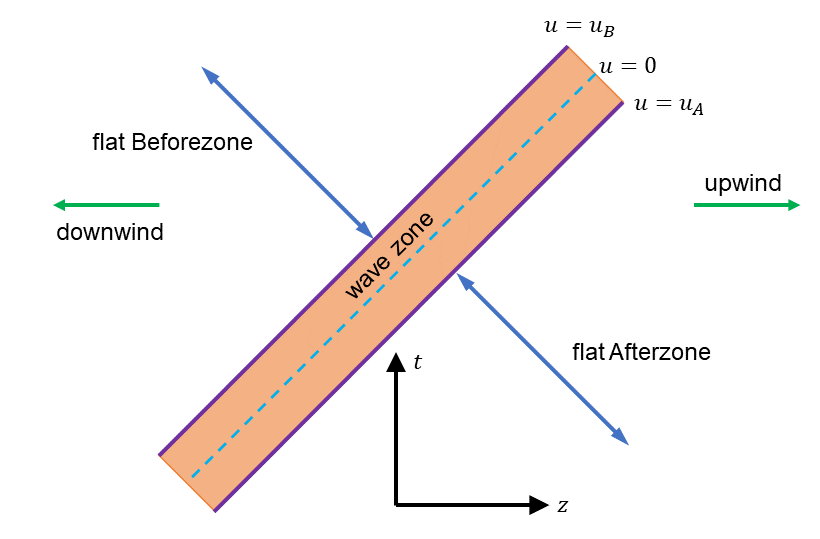}\vskip-3mm\caption{\textit{\small The sandwich wave
 propagates downwind. The space-time is flat both in the yet undisturbed Beforezone $U < U_b$
and  in the Afterzone $U > U_a $.
}
\label{BoPifig}
}
\end{figure}

We consider a plane \GW with $D=1$ transverse dimension whose metric is, in Brinkmann (B) coordinates $(X^\mu)=(X,U,V)$,
\beq
g_{\mu\nu}{dX^\mu}{dX^\nu}=
(dX)^2 + 2 dU dV - \half{\cA}(U)\,X^2 (dU)^2\,,
\label{Bmetric}
\eeq
\!where $X$ is space-like and $U,\, V$ are light-cone coordinates  \cite{Brink,Eisenhart,DBKP,DGH91}
\footnote{Our toy example is not a vacuum \GW because the coefficient of $dU^2$ is not traceless. It is a mere pp wawe - but this has no importance for us here.}.
The wave is assumed to be a sandwich wave  \ie, one whose profile $\cA(U)$ is zero in both the Beforezone $U<U_b$ and in the Afterzone $U>U_a$ and is non-vanishing only in a short Wavezone $U_b<U<U_a$ (see \cite{GibbHawk71,BoPi89,EZHRev} for the terminology, recalled in FIG.\ref{BoPifig}).

The geodesic motion is described by,
\begin{subequations}
\begin{align}
&\dfrac {d^2\!X}{dU^2} + \frac{1}{2}\cA X = 0\,,
\label{geoX}
\\[8pt]
&
\dfrac {d^2\!V}{dU^2}
 - \frac{1}{4}\dfrac{d\cA}{dU}(X)^2
 - \half\cA \frac{d(X^2)}{dU}
=0\,.
\label{geoV}
\end{align}
\label{Bgeoeqn}
\end{subequations}
\noindent
The spacelike coordinate $X$ is decoupled from the lightlike coordinate $V$ and the
projection  of a trajectory into transverse space
 is independent of $V$. Conversely, our  geodesic is a lift of  $X(U)$ determined by eqn. \eqref{geoX} alone, with $U$ viewed as Newtonian time \cite{DBKP,DGH91}.

 Equation \eqref{geoX} describes free motion
both in the Before and in the Afterzones where $\cA=0$, but {not} in the Wavezone  $\cA\neq0$, where  we have a Sturm-Liouville problem \cite{BoPi89,EZHRev}.
We first study the motion in the transverse space;
that of  $V$ is postponed to sec.\ref{LongiSec}.

Geodesics  admit a Jacobi invariant \footnote{
The Jacobi invariant is indeed a Casimir-invariant which has re-emerged recently  \cite{SchRev} in the E-D framework \cite{Eisenhart,DBKP,DGH91} for massive geodesics. It is reminiscent of Souriau's internal energy \cite{SouriauSSD}.},
 whose r\^ole will be highlighted in sec.\ref{massext}.
Discarding tachyons,
\beq
{\fm}^2 = -g_{\mu\nu}\dot{X}^{\mu}\dot{X}^{\nu} = \const \geq 0\,.
\label{Jacobiinv}
\eeq
For a massive  relativistic particle
initially at rest (as it is assumed in the study of the Memory Effect) the transverse initial conditions
are,
\beq
X(U_0) = X_0, \quad \dot{X}(U_0) = 0\,
\;\for U_0 \leq U_b\,.
\label{initcond}
\eeq
This paper is devoted to answer the question: \emph{When do we get pure displacement in the Afterzone~?} \ie,
\beq
\dot{X}(U)=0 \for  U > U_a\,.
\label{DMcond}
\eeq
We first consider massless particles,
$
{\fm}^2 =0\,;
$
the extension to the massive case ${\fm}^2>0$ will be discussed in sect. \ref{massext}.

\section{Gravitational Wave with Gaussian profile
}\label{GaussSec}

Our first example has a Gaussian profile. By rescaling the lightlike coordinate $U$  we can achieve that the wave has unit width,
\begin{equation}
\cA \equiv \cA^G(U) = \frac{k}{\sqrt{\pi}}e^{-U^{2}}\,,
\label{A0G}
\end{equation}%
and is normalized as
\beq
\int\!\cA(U)\,dU = k\,.
\label{Garea}
\eeq
The amplitude $k$ is thus the area below the profile.

\begin{figure}[ht]
\includegraphics[scale=.48]{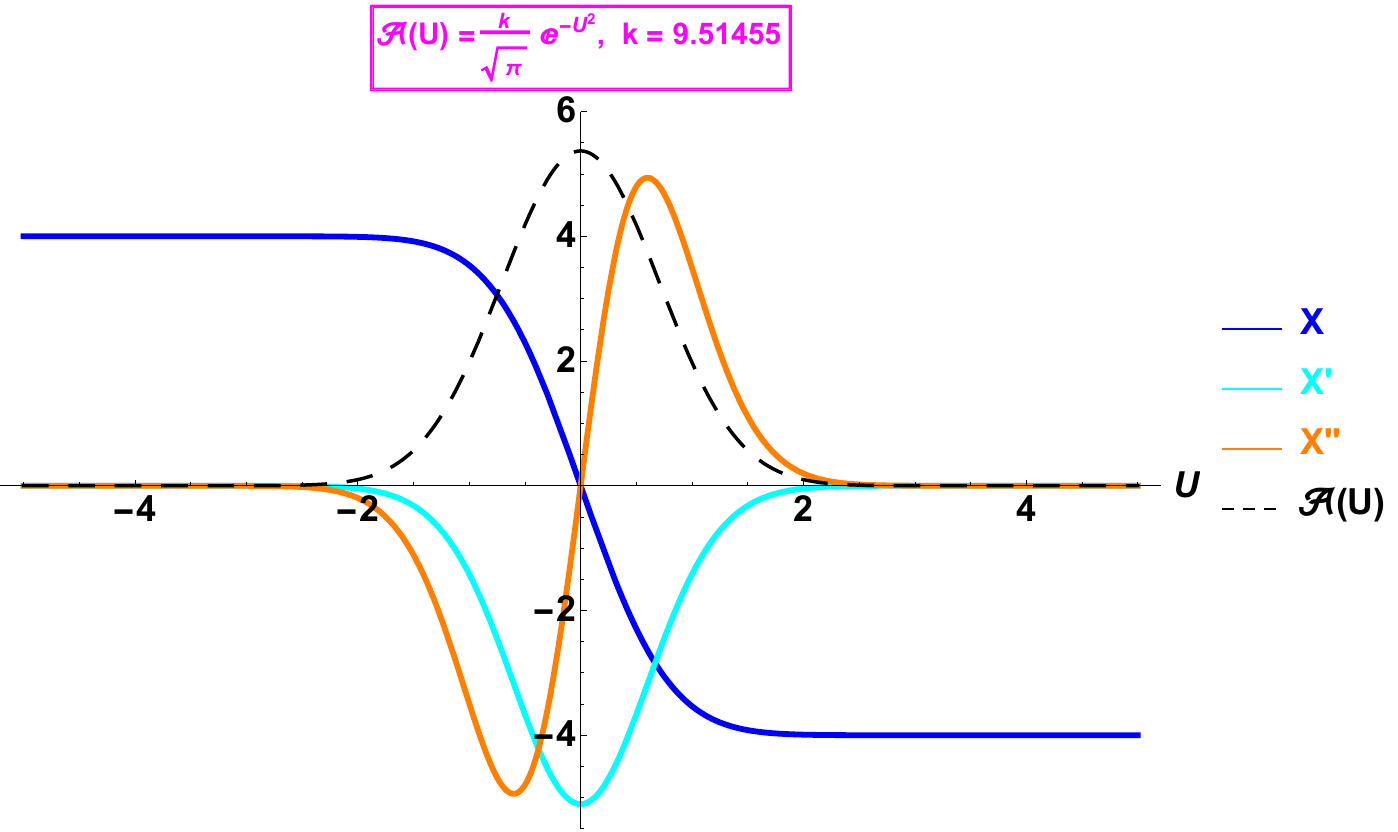}
\\
 \vskip-4mm
\caption{\textit{\small Fine-tuning the amplitude to $k=k_{crit} $ provides us with the ``half-wave displacement memory effect'' with ${\bf \magenta{m=1}}$  standing half-wave.
$\blue{X}: \blue{\bf trajectory}\,,
\;
\cyan{{dX}/{dU}}: \cyan{\text{\bf velocity}}\,,
\;
\orange{\bf d^2{X}/{dU^2}}: \orange{\text{\bf force}}$.
}
\label{Gauss-m1}
}
\end{figure}

Earlier work \cite{Ehlers,Hawking68,GibbHawk71,Sou73,GriPol,AiBalasin,BoPi89,ShortMemory,LongMemory,PolPer,SLheart,EZHRev}  indicated that after the passing of the \GW the particles fly apart with non-zero velocity : we have {VM}.
However now we show that \emph{fine-tuning the amplitude} $k$  can lead to (approximate) DM, as it will be illustrated by a series of figures.
Numerical investigations indicate, for example, that for  $k = k_{crit}=9.51455$
we get  a ``half-jump'', shown in FIG.\ref{Gauss-m1}. %

This ``miracle'' is explained, intuitively, by that at outside the (approximate) Wavezone $U_b < U < U_a$ both the \emph{velocity and the force} vanish, --- whereas the motion is governed by Newton's laws.

\begin{figure}[ht]
\includegraphics[scale=.34]{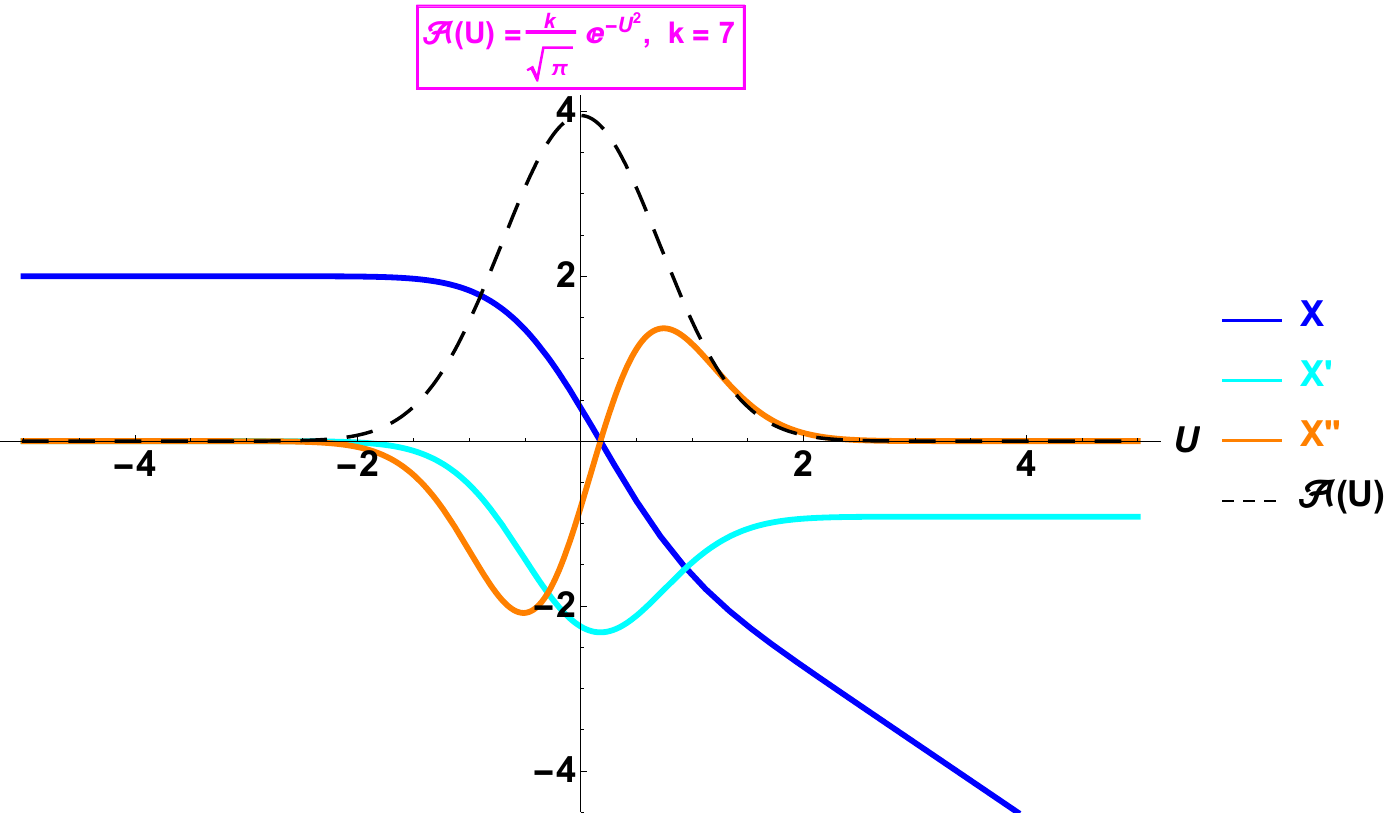}\;
\includegraphics[scale=.33]{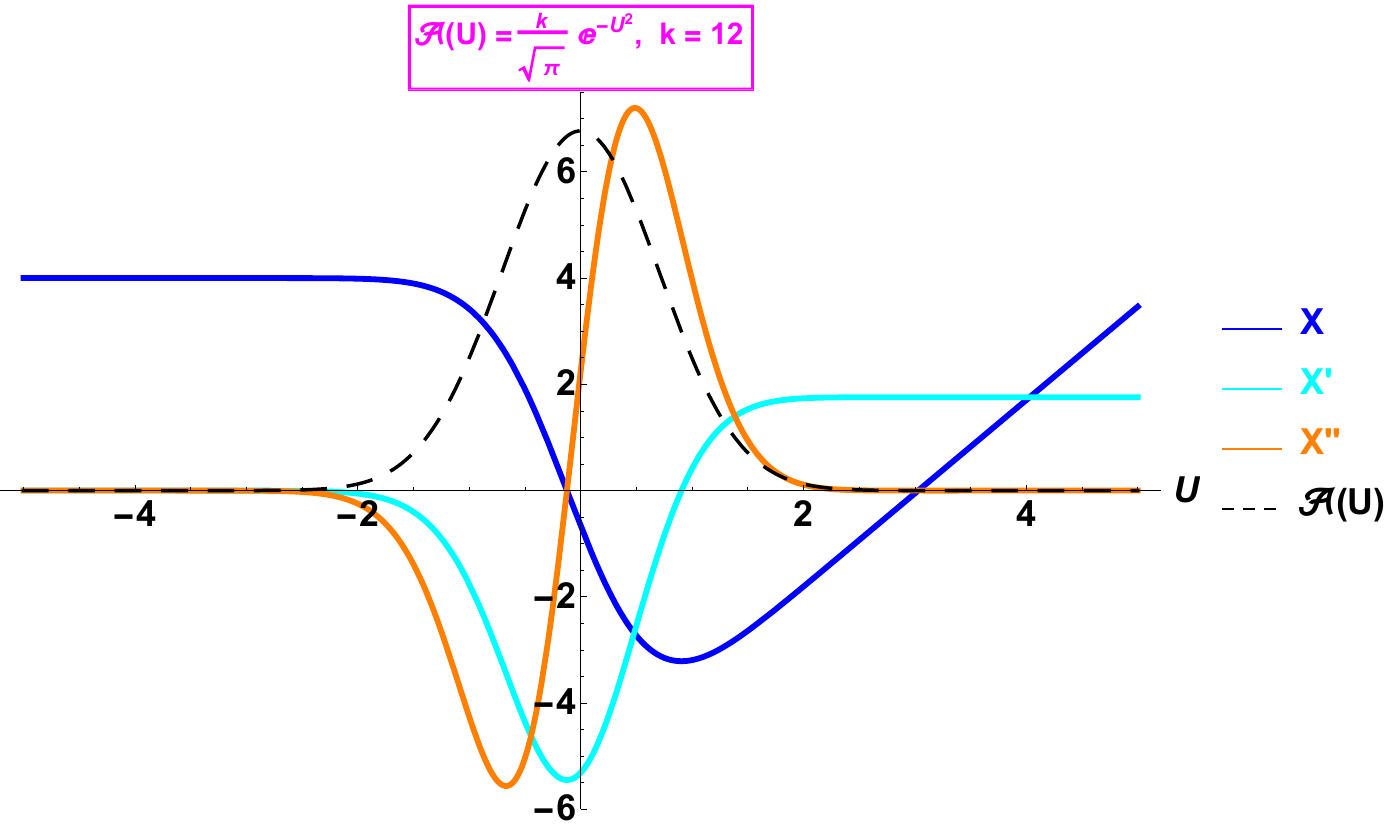}\\
\vskip-3mm\hskip-13mm
(a)\hskip75mm (b)
\vskip-3mm
\caption{\textit{\small (a) For  $k < k_{crit}$ the trajectory undershoots and (b) for $k > k_{crit}$ it overshoots before being straightened out.}
\label{knotcrit}
}
\end{figure}

FIG.\ref{knotcrit} confirms {VM} for $k\neq k_{crit}$, but show no {DM}~:
the velocity does not vanish (even approximately) in the Afterzone. For $k < k_{crit}$ the force falls off before the oscillator reaches its return point; for $k > k_{crit}$ it pulls instead the particle back after reaching it.

\begin{figure}[ht]\hskip-4mm
\includegraphics[scale=.35]{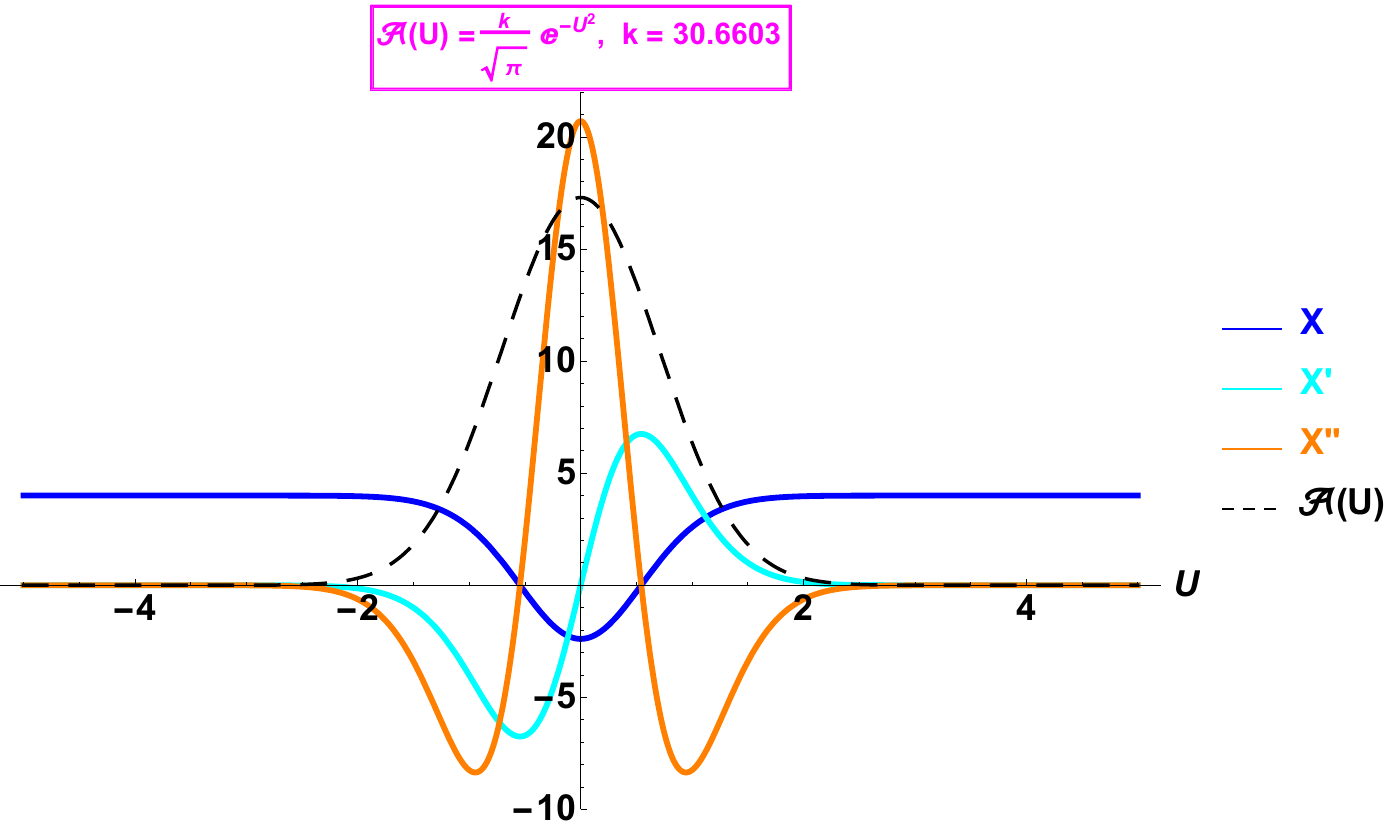}\;
\includegraphics[scale=.35]{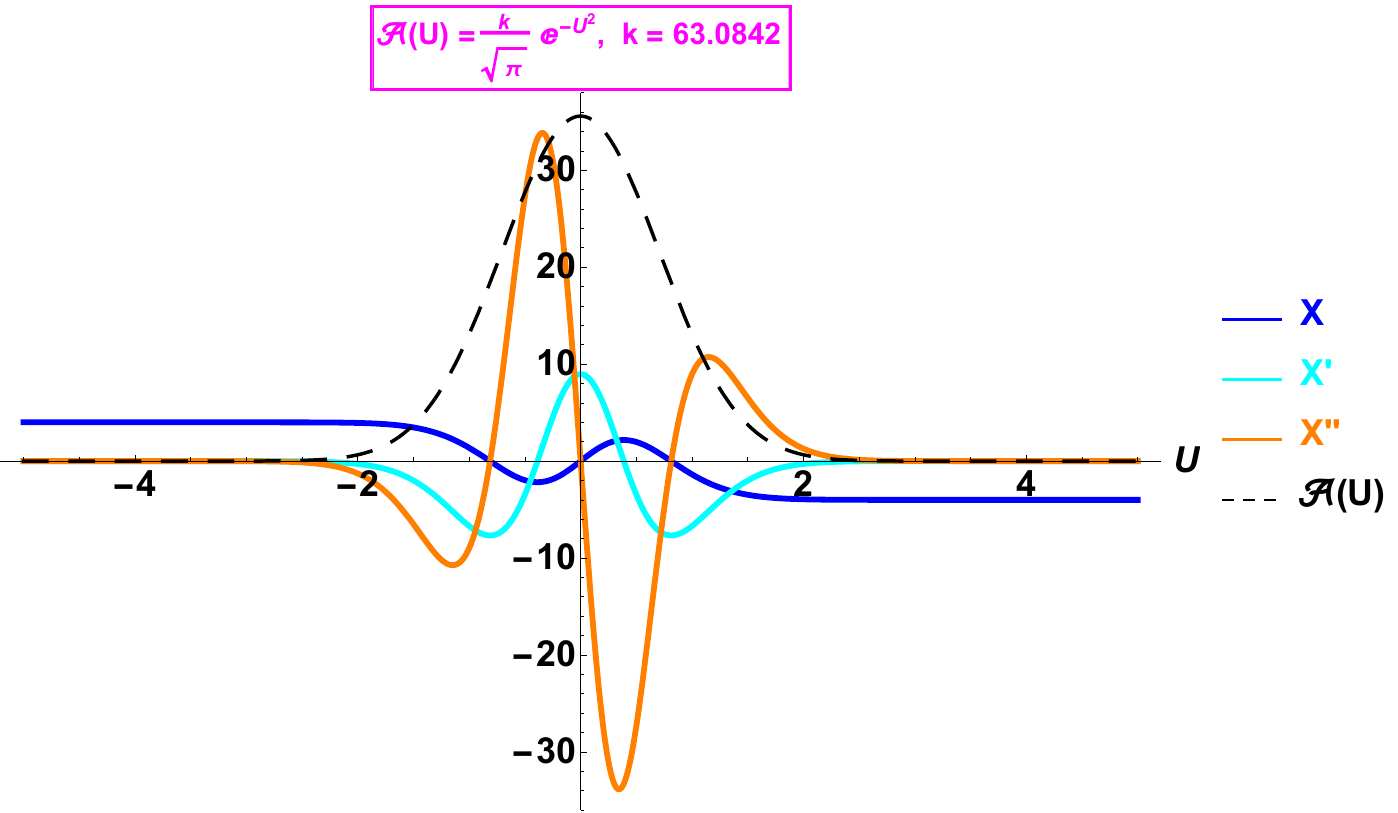}
\\
\vskip-4mm
\caption{\textit{\small Fine-tuning the amplitude yields  {DM} with ${\bf \magenta{m=2}}$ and
${\bf \magenta{m=3}}$ half-waves as trajectories.
NB: the plots have different scales.
}
\label{Gauss-m23}
}
\end{figure}

For $k_{crit}=k_1$ above we found precisely  trajectory consisting of $m=1$ one half-wave and one may wonder if {DM} with several half-waves can also be accommodated. The answer, obtained again by fine-tuning, says that {DM} can indeed arise with higher amplitudes when the \emph{Wavezone accommodates an integer number of half-waves}, as illustrated in FIG.  \ref{Gauss-m23}.

Emboldened by this success, further fine-tuning yields {DM} for other magic amplitudes,
\beq
k_1 \approx 9.5,\,   m=1, \;\;\;
k_2 \approx 30.7,\,  m=2, \;\;\;
k_3 \approx  63.1,\, m=3, \;\;\;
k_4 \approx  106.7,\,m=4, \, \dots
\label{G-magicm}
\eeq
The outgoing position depends on the parity of $m$ :
\beq
X_{out}=(-1)^mX_{in}\,.
\label{Xinout}
\eeq

\begin{figure}[ht]
\includegraphics[scale=0.25]{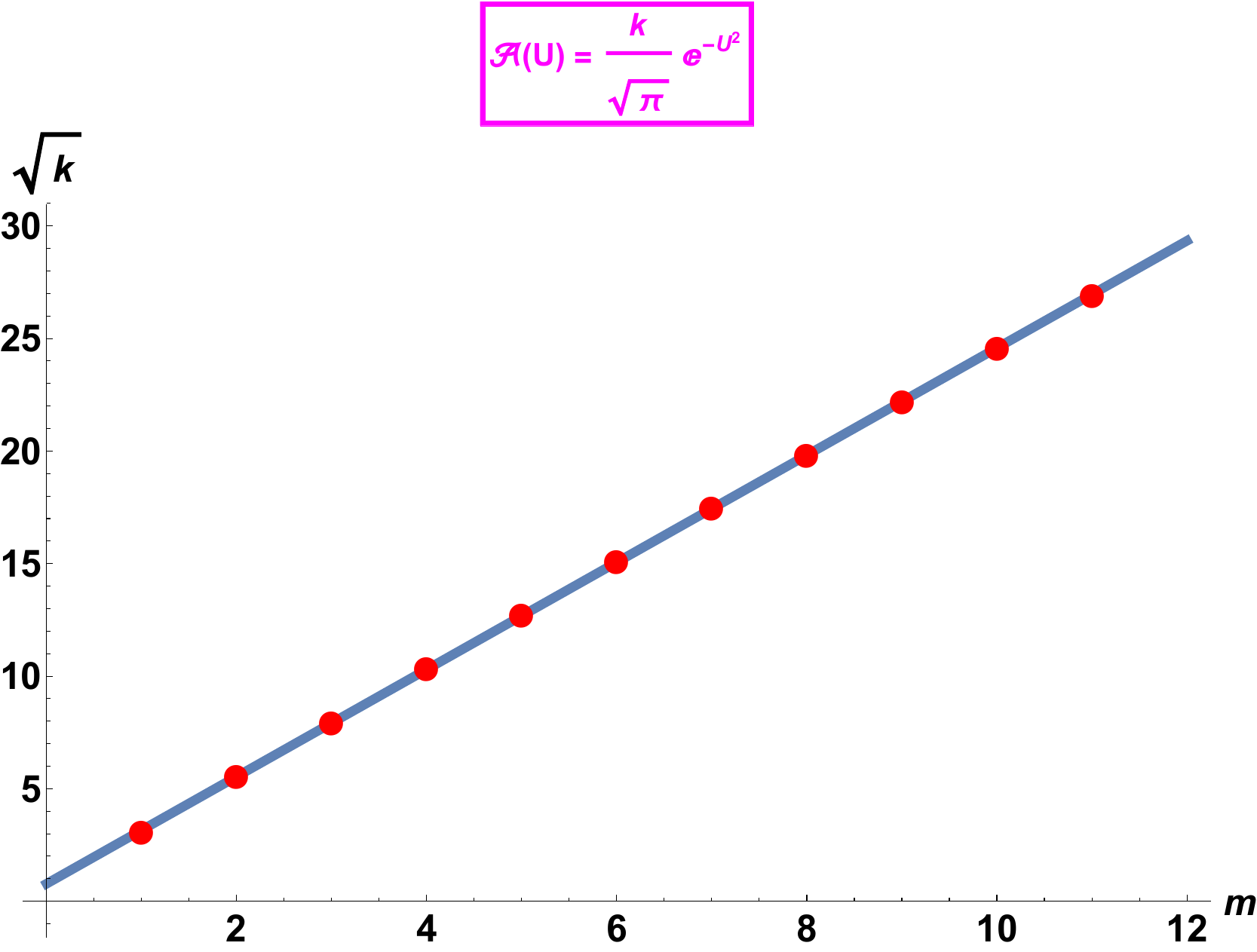}\vskip-3mm
\caption{\textit{\small The relation between ${\bf \magenta{m}}$, the number of half-waves in the trajectory in the Wavezone  and $\sqrt{k_{crit}}$ for {DM} is approximately linear.}
\label{Gauss-km}
}
\end{figure}

Higher wave number requires higher amplitude.
The relation between $\sqrt{k}$ and $m$,
 depicted in FIG.\ref{Gauss-km}, is approximately linear,
\beq
 \sqrt{k_m} \approx 0.78 + 2.38 m\,.
 \label{Gkmrel}
\eeq\vskip-3mm
\goodbreak

Intuitively, the coefficient of $dU^2$ in \eqref{Bmetric} is the potential of an oscillator and
the particle initially at rest at $X_0\neq0$
 is pulled towards the origin, picking up some speed. However passing $X=0$ the force changes direction and starts to reduce the velocity.
If the profile was $U$-independent, --- it would be a genuine harmonic oscillator --- then the particle would  oscillate between $X_0$ and $-X_0$ forever. However for sandwich waves with bell-shaped potentials the pull progressively falls off with increasing $U$ and after a while  the motion becomes free.
If the residual velocity is non-zero, then we have {VM} as in FIG. \ref{knotcrit}. However if the velocity is zero which happens after an integer number  of half-oscillations, then the particle stops, --- and  then it does not restart anymore by virtue of Newton's laws which are valid in the Afterzone: we get {DM}.

 Higher amplitude $k$ means stronger force which then requires more back-and-forth oscillations i.e. a larger $m$ before stopping.

These arguments are largely independent of the concrete profile as long as it is roughly bell-shaped, as it will be illustrated on the analytic example discussed below in sec.\ref{PTSec}.

\section{Gravitational Wave with P\"oschl-Teller profile
}\label{PTSec}

\begin{figure}[ht]
\includegraphics[scale=.35]{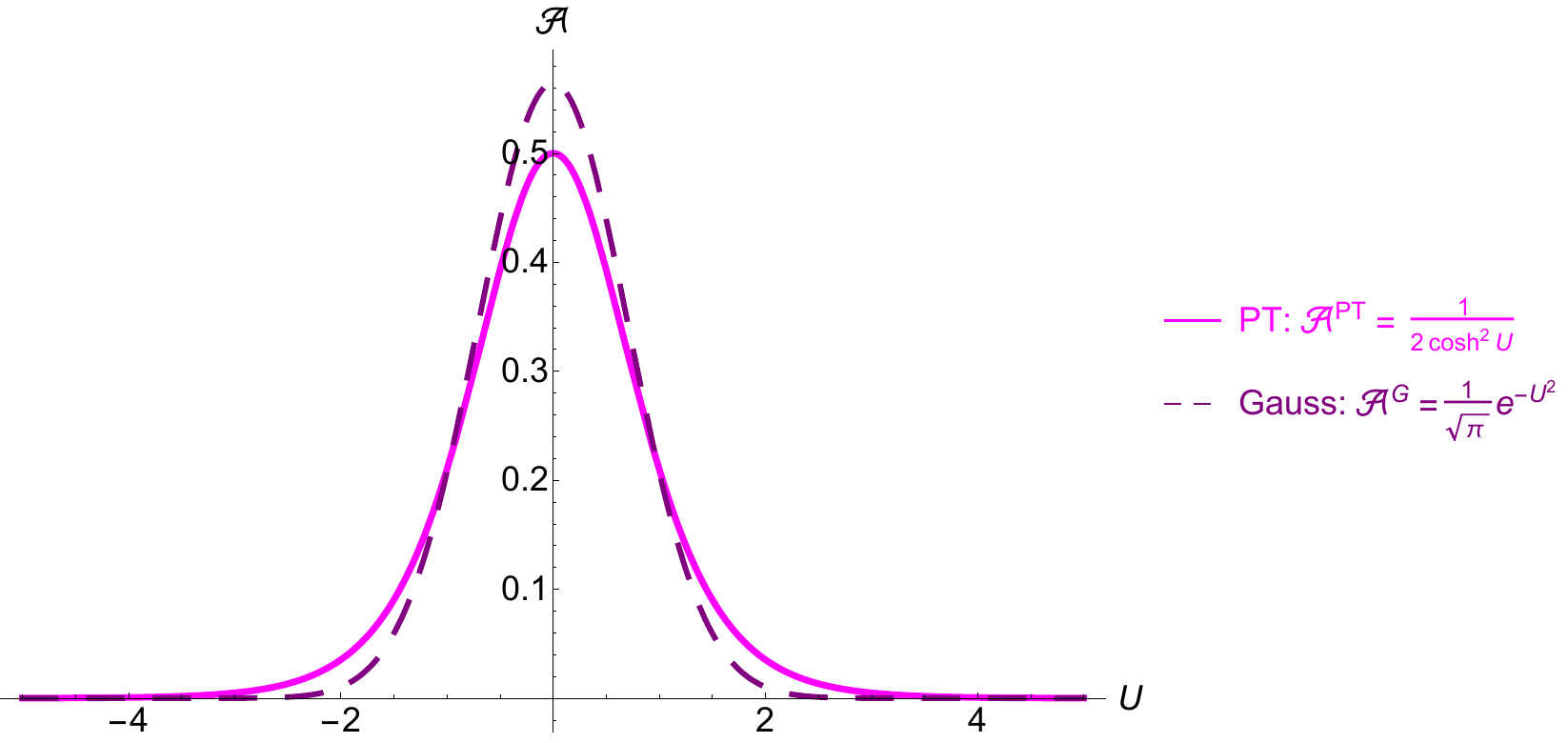}
 \vskip-3mm\caption{\textit{\small
The \purple{Gaussian bell} (dashed) can be approximated by the
\magenta{P\"oschl-Teller} potential \eqref{PTPot} (solid line), which admits analytic solutions. The parameters were chosen so that the area below both profiles be identical and equal to $k$.}
\label{PT-pot}
}
\end{figure}

\begin{figure}[ht]
\includegraphics[scale=.48]{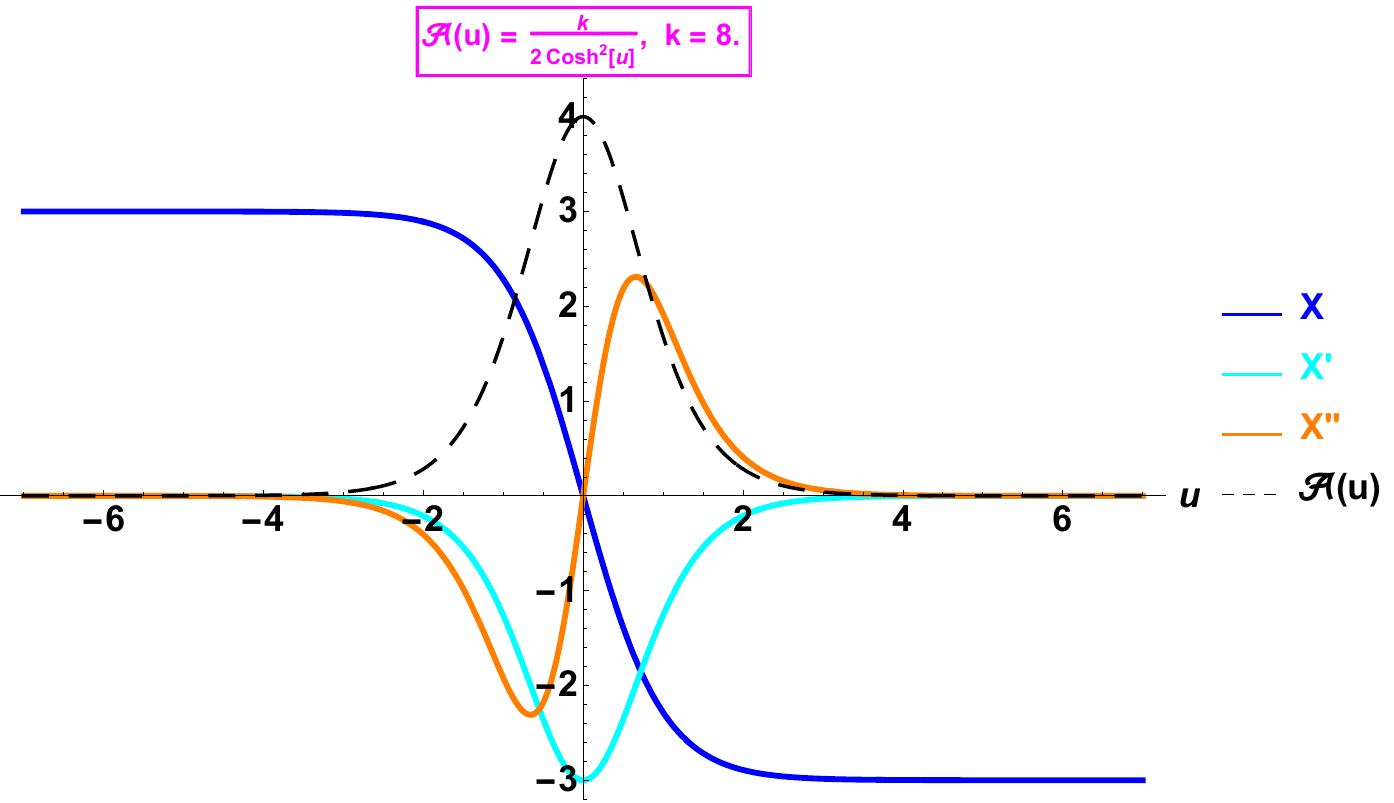}\;
\\
\vskip-3mm\caption{\textit{\small For the \PT
profile with $k_{crit}=k_1=8$ i.e. ${\bf \magenta{m=1}}$,  the transverse trajectory is consistent with DM (to be compared with FIG.\ref{Gauss-m1}).}
\label{PT-m1}
}
\end{figure}

\begin{figure}[ht]
\includegraphics[scale=.34]{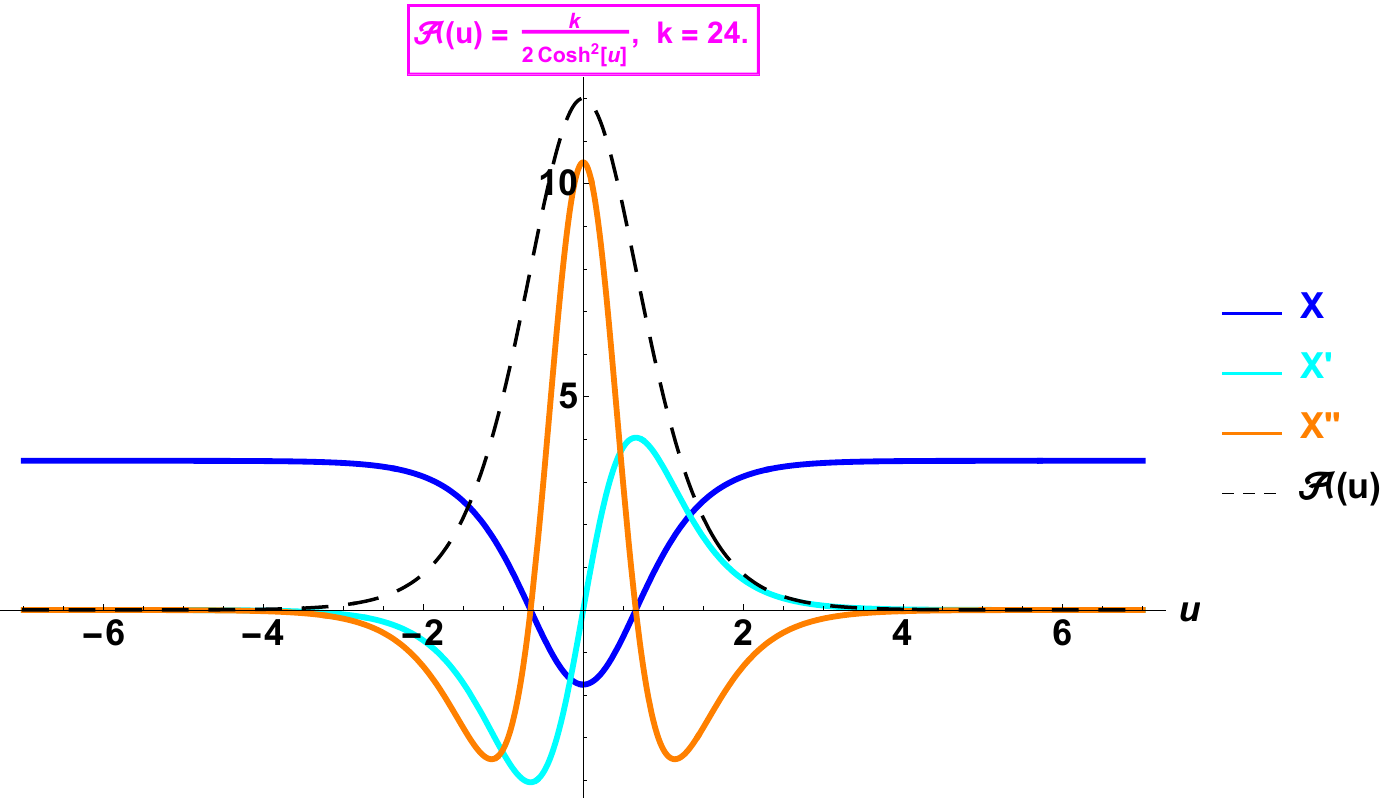}\;
\includegraphics[scale=.345]{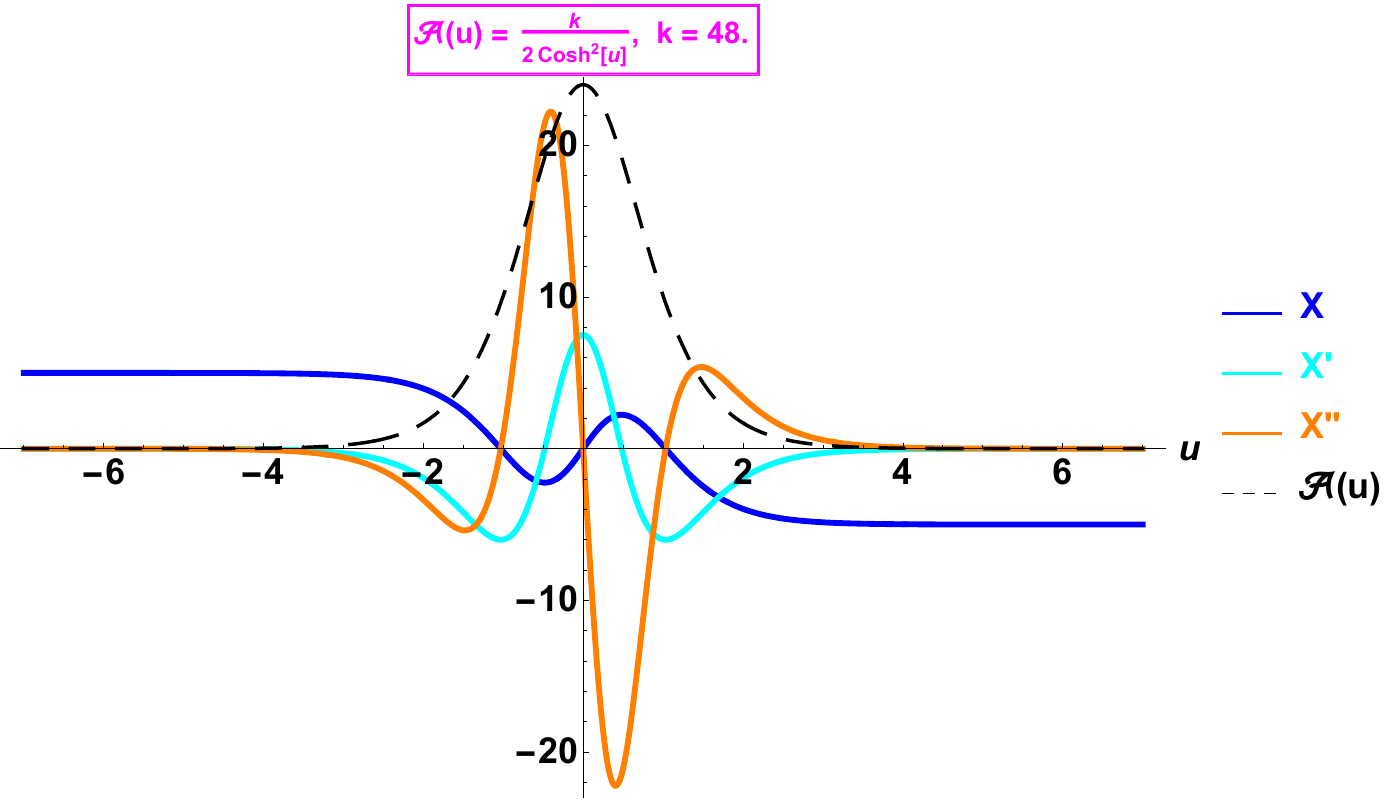}\\
\vskip-2mm\hskip-14mm
(a)\hskip78mm (b)
\vskip-3mm\caption{\textit{\small The trajectory \eqref{PT-XmU} for the \PT
profile with (a) $k_{2}=$ 24 and (b) $k_{3}=$ 48 have standing waves with wave numbers ${\bf \magenta{m=2}}$ resp.
${\bf \magenta{m=3}}$, cf. FIG.\ref{Gauss-m23}.}
\label{PT-m23}
}
\end{figure}

\begin{figure}[ht]
\includegraphics[scale=.4]{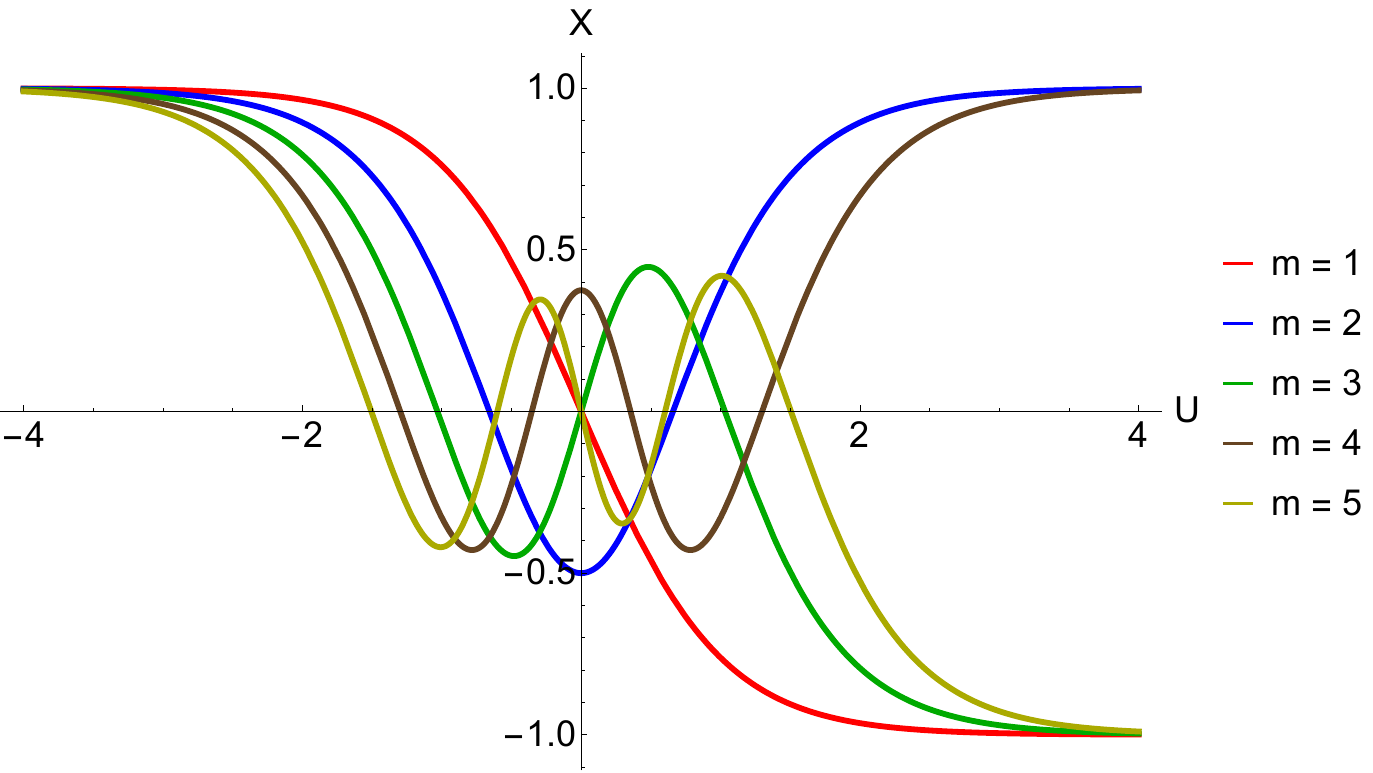}\vskip-3mm\caption{\textit{\small
Transverse trajectories for the \PT profile with \magenta{$m= 1, \dots, 5$}.
}
\label{PT-m15}
}
\end{figure}

\begin{figure}[ht]
\includegraphics[scale=0.3]{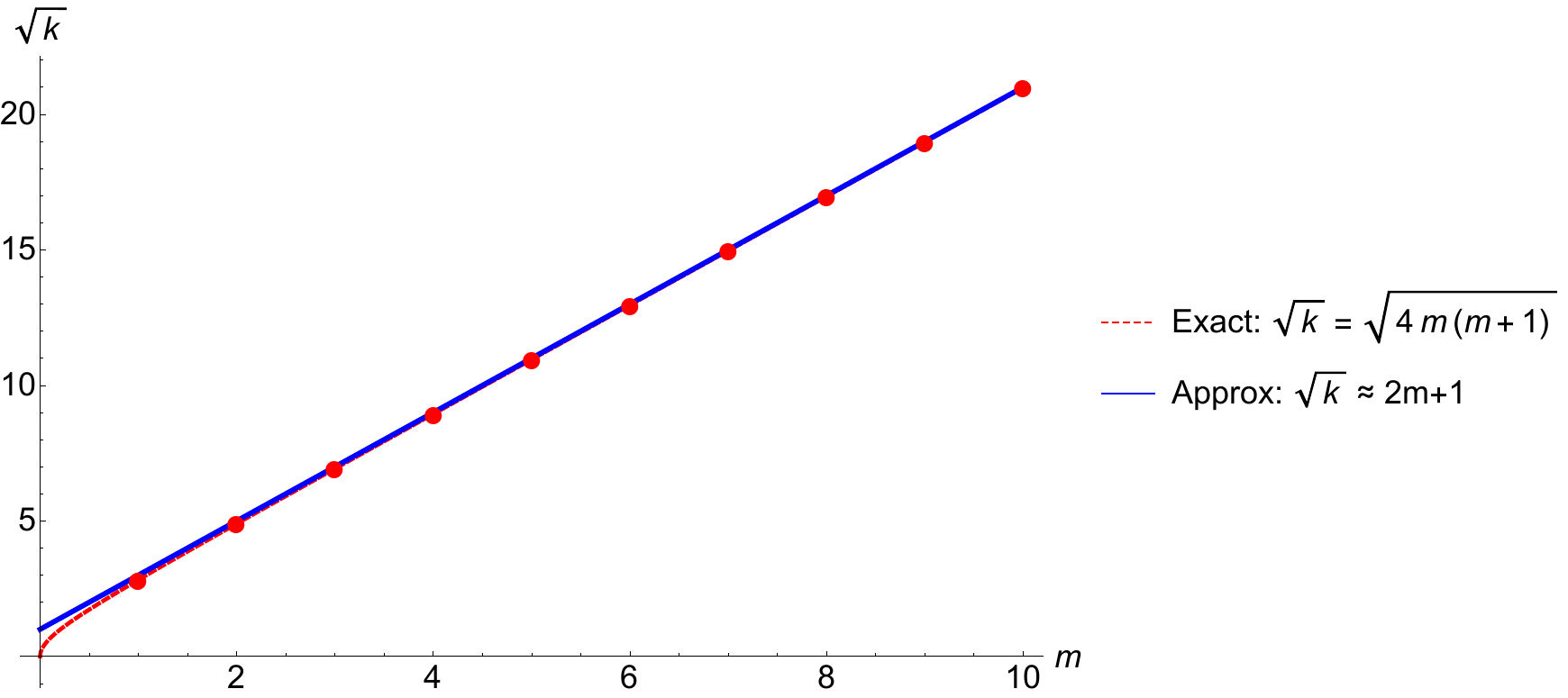}\vskip-3mm
\caption{\textit{\small The relation between  the number of half-wave trajectories in the Wavezone, ${\bf \magenta{m}}$, and $\sqrt{k_{crit}}$ for {DM} is approximately linear, as it is for the Gaussian in FIG.\ref{Gauss-km}.}
\label{PT-km}
}
\end{figure}

For Gaussian profile the Sturm-Liouville equation \eqref{geoX} has no analytic solution. However the shape of $\cA$ is strongly reminiscent of the [symmetric part of the]  P\"oschl-Teller (PT) potential~\cite{PTeller}, considered, independently, also in \cite{Chakraborty:2019yxn} and in \cite{ElBaZh},
\beq
\cA^{PT}(U) = \dfrac{k}{2\cosh^2 U}\,,
\label{PTPot}
\eeq
depicted in FIG.\ref{PT-pot} .
%
$\cA^{PT}$ is normalized as the Gaussian, \eqref{A0G},
$
\int\!\cA^{PT}(U)dU = k\,,
$
cf. \eqref{Garea}. It has the advantage
 that the trajectories can be found {analytically}. Writing
\beq
k = k_m = 4 m(m+1),
\label{PTkm}
\eeq
where $m$ is {\rm a priori} a real number, eqn. \eqref{geoX} becomes that of a damped oscillator whose frequency
$\omega^2=\frac{m\left(m+1\right)}{\cosh^{2}U}$
decreases with $U$,
\begin{equation}\bigbox{
\dfrac{d^{2}X}{dU^2}+\frac{m\left(m+1\right)}{\cosh^{2}U}\,X=0\,.}
\label{PTtraj}
\end{equation}%
The initial conditions for a particle at rest before the burst arrives are,
\beq
X(U=-\infty)=X_0, \aand \dot X(U=-\infty)=0\,. %
\label{Xinit}
\eeq
Putting $t=\tanh (U)$ into \eqref{PTtraj} the Legendre equation is obtained,
\begin{equation}
\left(1-t^{2}\right) \frac{d^{2}X}{dt^{2}}-2t\frac{dX}{dt}+m\left(m+1\right) X=0\,.
\label{Legendreeq}
\end{equation}%
Then {DM} means that $X(U)$  \emph{tend to a constant} for $U\to \infty$ which amounts to requiring that
 the solution of \eqref{Legendreeq} should extend to $t=\pm1$ which in turn implies  that  \emph{$m$ must be a positive integer}, and the solution becomes proportional to a \emph{Legendre polynomial},
\beq
X(U) =  X_m(U) = (-1)^m\,P_m(\tanh U)\,X_0,
\qquad
m = 1,\, 2,\, \dots\,,
\label{PT-XmU}
\eeq
shown in FIGs.
\ref{PT-m1}-\ref{PT-m23}-\ref{PT-m15} should be compared with \eqref{G-magicm} and with
FIGs.\ref{Gauss-m1} and \ref{Gauss-m23} for the Gaussian).
The trajectories \eqref{PT-XmU} are composed of ${m}$ half-waves, as for the Gaussian. The \PT counterpart of the $k_{crit} \Leftrightarrow m$ relation \eqref{Gkmrel},
\beq
k_m = 4m(m+1) \Rarrow \sqrt{k_{m}} \approx 2m + 1\,,
\label{PTkmrel}
\eeq
reminiscent of \eqref{G-magicm} is
shown in FIG.\ref{PT-km} to be compared with
FIG.\ref{Gauss-km}.

\begin{figure}[ht]
\includegraphics[scale=.45]{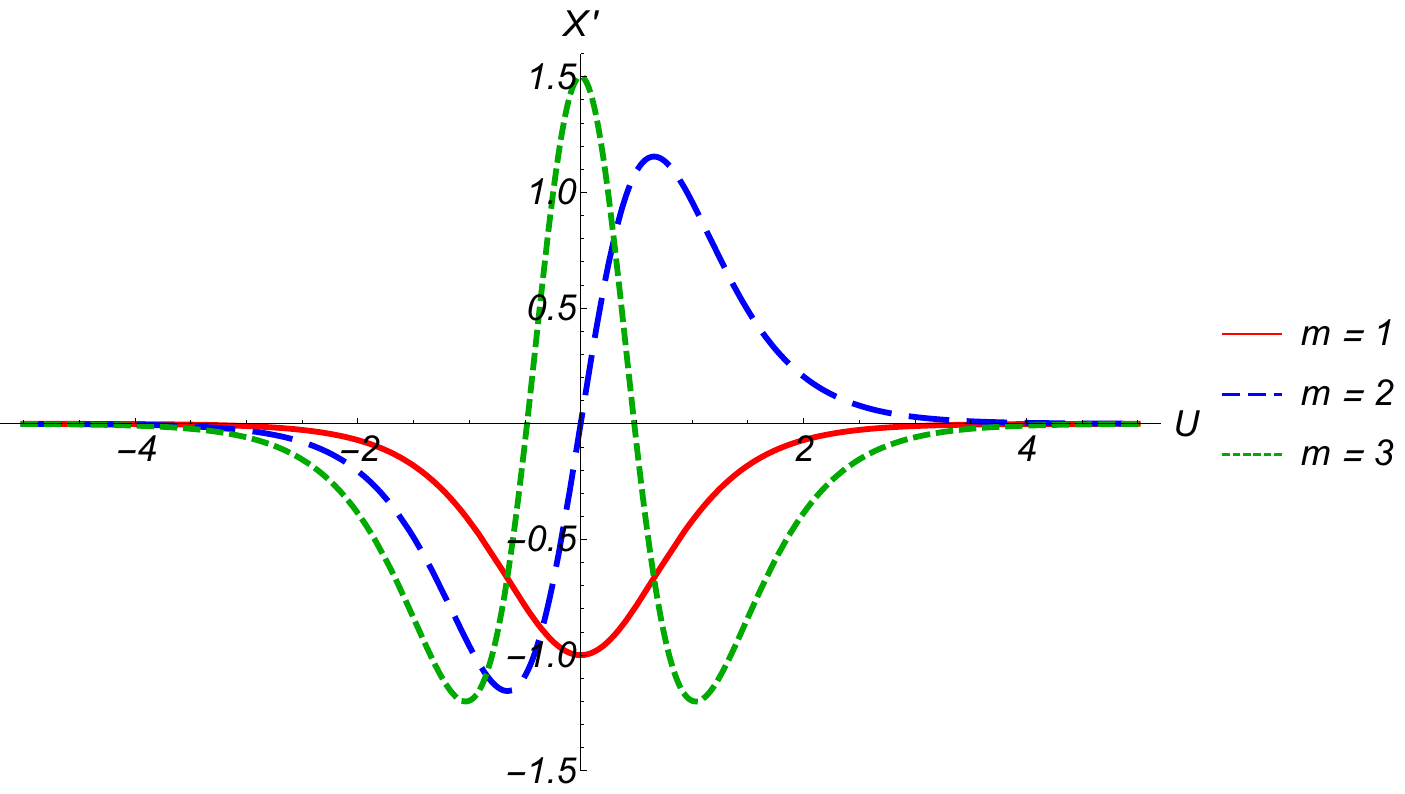}\\
\vskip-5mm\caption{\textit{\small The transverse velocities
 for the  \PT profile \eqref{PTPot}-\eqref{PTkm},
 for DM amplitudes $k=k_{m}$,
 shown for \magenta{${\bf m}$} = \red{${\bf 1}$}, \blue{${\bf 2}$}, \dgreen{${\bf 3}$}.
}
\label{PT-velocity-m123}
}
\end{figure}

For $k\neq k_{crit}$ we get again {VM} but no {DM}, as it could be illustrated by plots similar to those in FIG.\ref{knotcrit}.
It is instructive to plot also the velocities,
  FIG.\ref{PT-velocity-m123}, which confirms that
 the number of half-waves is equal to the number of zeros of $dX_m/dU$ in the wave zone \footnote{
An analytic proof can be found by using the properties of the Legendre polynomials, and in particular
$
P_n'(x) = n\frac{P_{n-1}(x) - x P_n(x)}{1-x^2}
$.}.

We conclude this section by pointing out a curious relation with quantum mechanics.
Let us indeed consider the one dimensional time-independent Schr\"odinger equation for the \PT potential, %
\begin{equation}
-\frac{\hbar^{2}}{2M}\frac{d^{2}\psi}{dx^{2}}
-\frac{\hbar^{2}}{2M}\frac{m(m+1)}{\cosh^{2}x}\, \psi = E\psi
\end{equation}
where the negative sign was chosen to get bound states. %
Redefining the energy as $\varepsilon =\frac{2M}{\hbar^{2}}E$ then yields
\begin{equation}
-\frac{d^{2}\psi}{dx^{2}}-\frac{m(m+1)}{\cosh^{2}x}\,\psi
=\varepsilon\, \psi\,.
\label{QMenergyeq}
\end{equation}%
Substituting here
$
\psi \to X,
\,
x \to U
$
and putting $\varepsilon=0$
we recognize our equation  \eqref{PTtraj}~: our {DM} trajectories correspond to \emph{zero-energy bound states} of the quantum problem, shedding some light at the curious quantization of wave numbers, $m$. Such zero-energy solutions are discarded in quantum mechanics because they are not normalizable -- but this condition is not required for our trajectories and  the zero-energy solutions in \eqref{PT-XmU} which are thus admissible.

\section{Longitudinal motion}\label{LongiSec}

Returning to the general setting, now we complete our study of the ${\fm}=0$ case by extending DM to the ``vertical'' component $V(U)$.
Let us recall that eqn. \eqref{geoV} is obtained by lifting the transversal trajectory $X(U)$ to Bargmann space,
\beq
\hat{V}(U)=
 V_0 - \cI(U)\,,
\qquad \cI(U) = \int_{-\infty}^U\!\!{\cL}_{NR}\,du\,,
\label{nullV}
\eeq
where ${\cL}_{NR}$ is the Lagrangian of a non-relativistic particle in $1+1$ dimensions which moves in a possibly time-dependent oscillator potential;
$\cI(U)$ is  the classical Hamiltonian action of the underlying NR model calculated along $X(U)$. We have chosen $M=1$ for simplicity.

DM in transverse space requires the initial and final conditions
\begin{equation}
X^{\prime}(U=-\infty)=0=X^{\prime}(U=\infty)\,.
\label{DMbound}
\end{equation}
The integral  $\cI$ in \eqref{nullV} makes the theory  {\rm a priori} non-local. However using
 the equations of  motion \eqref{geoX}  a calculation shows that along a trajectory,
\begin{eqnarray*}
\int\! \cL_{NR}du&=&\frac{1}{2}\int X^{\prime 2}du-\frac{1}{4}\int{\cA}X^{2}du =
\frac{1}{2}\int X^{\prime}dX-\frac{1}{4}\int X\left(-2X^{\prime \prime
}\right) du
\\[6pt]
&=&\frac{1}{2}XX^{\prime}\Big|_{-\infty }^{\infty }-\frac{1}{2}\int XX^{\prime
\prime}du+\frac{1}{2}\int XX^{\prime \prime}du
=\frac{1}{2}XX^{\prime }\Big|_{-\infty}^{\;\infty}
\end{eqnarray*}
so that $\cI$ is in fact local.
Moreover, for the {DM boundary conditions}  \eqref{DMbound} it vanishes,
\beq\medbox{
\cI =
\int_{-\infty}^{\;U}\!\cL_{NR}\,du=0\;
\for\; U > U_a\,.
}
\label{intL0}
\eeq
The (possibly) non-local term is thus eliminated, leaving us with,
\beq\medbox{
\hat{V}(U)=V_0
\quad \text{\small both for}\quad U < U_b \aand
U > U_a\,}
\label{nullVV0}
\eeq
which extends, for $\fm=0$, DM from the transverse coordinate to $V$ (with no $V$-displacement as all as a bonus).

\begin{figure}[ht]
\hskip-2mm\includegraphics[scale=.31]{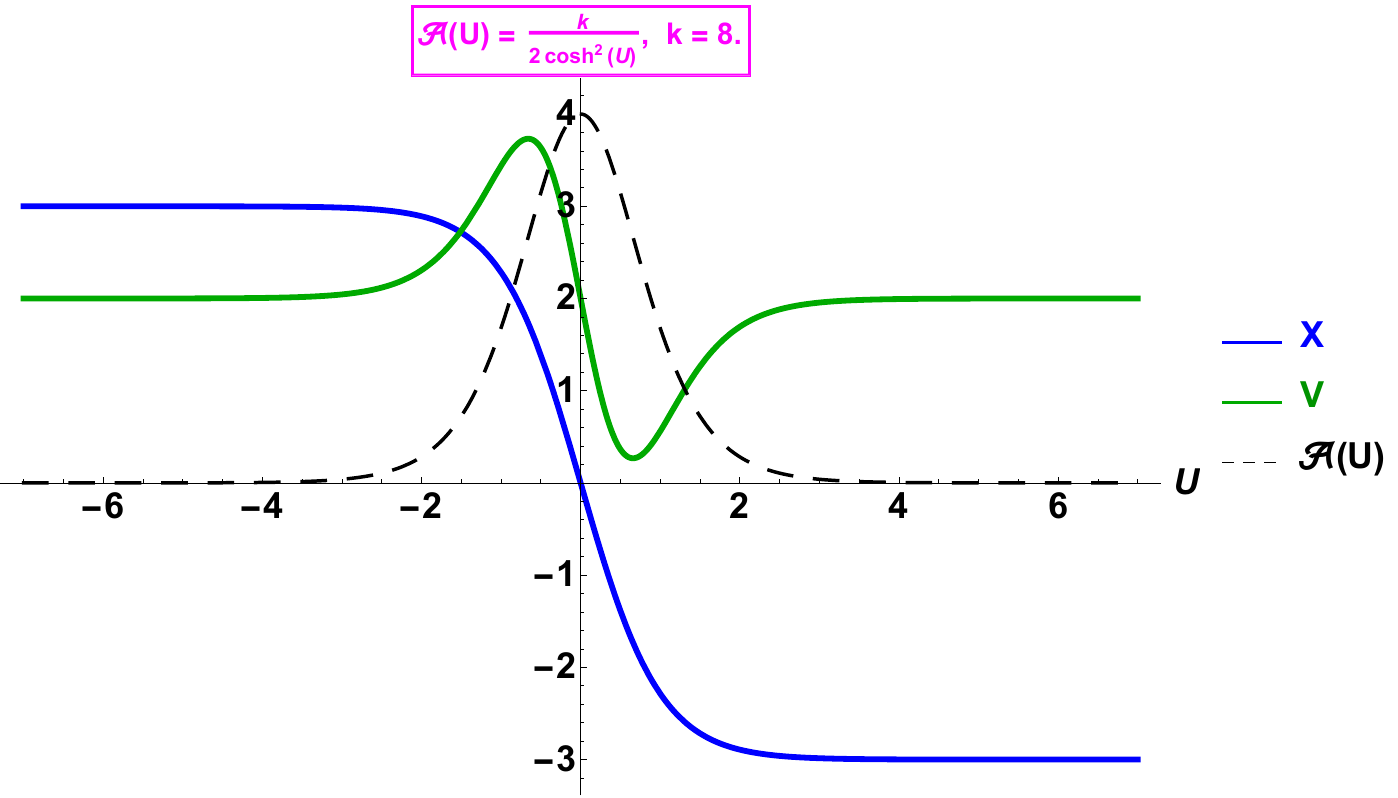}
\includegraphics[scale=.355]{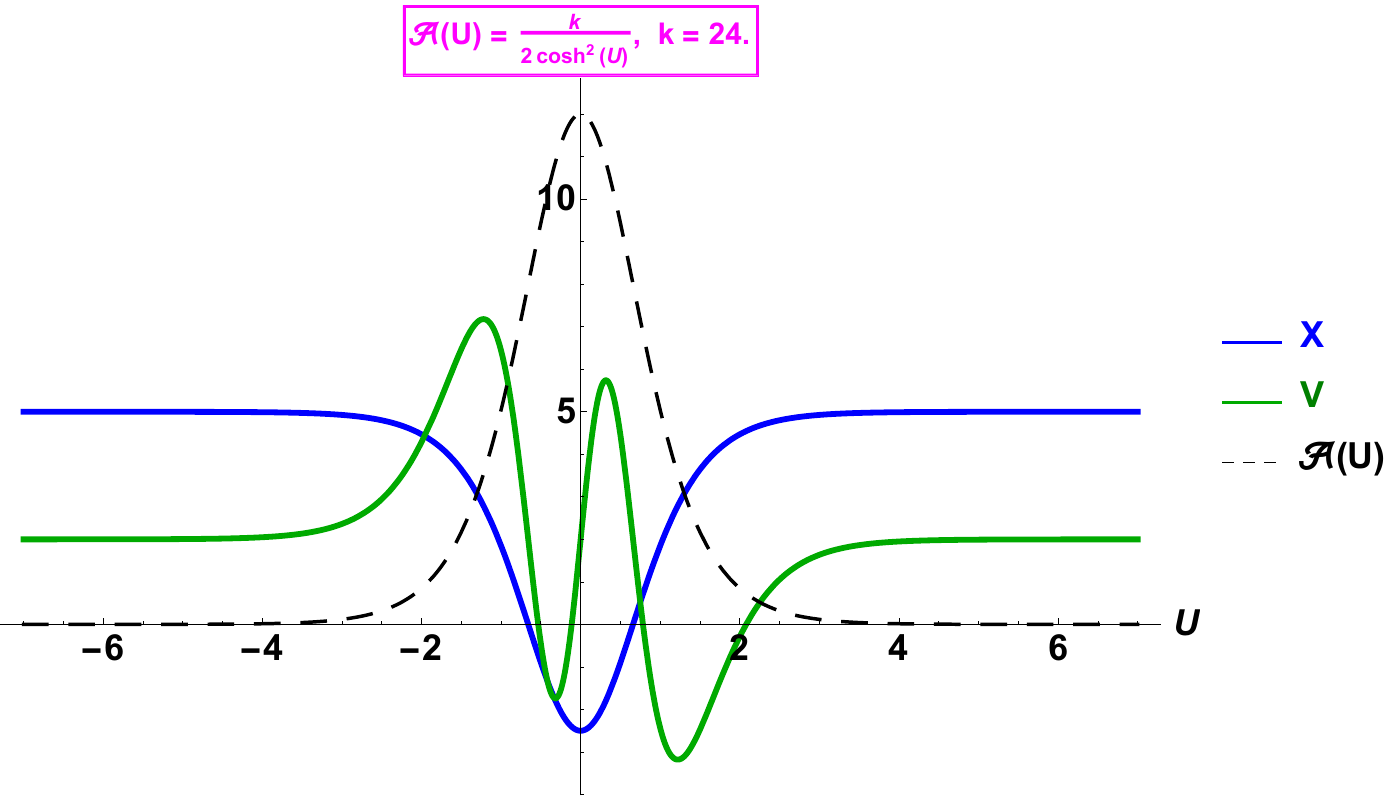}\\
\vskip-2mm\hskip-20mm
(a)\hskip74mm (b)
\vskip-3mm\caption{\textit{\small For the \PT
profile with $k_{crit}=k_m$ shown for
${\bf \magenta{m=1,\,2}}$
both the \blue{\bf transverse, $X$}, and the \dgreen{\bf vertical, $V$}, trajectories behave, in the \emph{massless} case $\fm=0$, consistently with DM. NB:  the  scales for (a) and (b) are different.}
\label{PT-XV}
}
\end{figure}

 Non-trivial motion arises only in the Wavezone, as shown in FIG.\ref{PT-XV}.

Similar plots could be obtained (numerically) for the Gaussian.

We underline that \eqref{intL0} is valid only when the domain of integration contains the  Wave zone.
 In the \PT case, for example,
 \beq
\left\{\barraynb{llllc}
m = 1 &\qquad  &\hat{V}(U) = &V_0 & -\dfrac {1}{2} \dfrac{\sinh (U)}{\cosh^3(U)}\, {X_0^2}\,
\\[18pt]
m = 2 &\qquad  &\hat{V}(U) = &V_0& \quad
-\left(\dfrac{3}{2}\dfrac {\sinh(U)}{\cosh^3(U)} - \dfrac  9 4 \dfrac {\sinh(U)}{\cosh^5(U)}\right)X_0^2
\,.
\earraynb\right.
\label{m1m2V}
\eeq
However the non-trivial terms fall off to zero outside the (approximate) Wavezone, as illustrated in  FIG.\ref{PT-XV}.

Our plots indicate that while DM requires that the transverse trajectory be composed of $m$ half-waves, the ``vertical'' $V$ has $m$ full waves.

\section{Extension to the massive case}\label{massext}

Requiring zero incoming total velocity, \eqref{initcond}, is however \emph{unphysical} for a massless particle: one can not stop a photon. Hence the importance of extending  our investigations to particles with nonzero Jacobi invariant, $\fm\neq0$, which do not follow null geodesics \cite{Eisenhart, EDAHKZ, SchRev}.

The $V$-equation \eqref{geoV} can actually be solved for $\fm^{2} \geq 0$~\footnote{We are grateful to J. Balog and G. Junker for calling our attention at this point.}. To this end we rewrite it by inserting  $X(U)$  from \eqref{geoX}, as
\beq
\dfrac {\,d^2V}{dU^2} = -\dfrac {d}{dU}
\underbrace{\left(
 \dfrac 1 2 \big(\dfrac {\,dX}{dU}\big)^2 - \dfrac 1 4 {\cal A} X^2\right)}_{\cL_{NR}}\,,
\eeq
where in the bracket we recognize again the \emph{non-relativistic Lagrangian}, $\cL_{NR}$.
Integrating twice then yields,  consistently with previous results, \cite{Eisenhart,EDAHKZ,SchRev},
\beq
V  = \left\{V_0 - \int\!\cL_{NR}\, dU\right\}\,  + V_c\,U
= \hat{V}+ V_c\, U\,,
\label{VUint}
\eeq
where $V_0$ and $V_c$ are integration constants.
The meaning of $V_0$ is obvious, however what is the physical role of $V_c$ ? To this end,
we first emphasise that our theory admits in fact \emph{two, different mass} parameters. One of them is the relativistic mass, $\fm$, defined by the Jacobi invariant \eqref{Jacobiinv}.
 The other one we denote by $M$ is the conserved quantity associated with the ``vertical'' Killing vector $\p_V$ which plays, in the E-D framework, the role of mass in the underlying NR theory \cite{Eisenhart,DBKP,DGH91}.

 We note   that for an affine parameter $\lambda$ the Jacobi invariant is  $g_{\mu\nu}dx^{\mu}dx^{\nu }=-\fm^{2}d\lambda^{2}$ and then switching to $U$ implies, after rearrangement,
\beq
\frac{dV}{dU}=-\frac{1}{2}\left(\frac{dX}{dU}\right)^{2}
+\frac{1}{4}{\cA}X^{2}-\frac{1}{2}\frac{\fm^{2}}{M^{2}}
=-\cL_{NR}-\frac{1}{2}\frac{\fm^{2}}{M^{2}}\,,
\label{dVdU}
\eeq
whose integration over a large-enough domain then yields  \eqref{VUint} with
\beq
 V_c= - \frac{\fm^{2}}{2M^{2}}\,.
\label{VcmM}
\eeq

At last, the {DM boundary conditions} for the transverse motion $X(U)$ in \eqref{DMbound} imply that the integral term $\displaystyle\int\!{\cL}_{NR}dU$ (which is independent from $\fm$)  drops out, cf. \eqref{intL0}, and thus  we end up, outside the wavezone, with the generalization of \eqref{nullVV0} to the massive case,
\beq
\bigbox{
V \equiv V_{\fm}(U)=V_{0} - \frac{1}{2}\frac{\fm^{2}}{M^{2}}\,U \,.}
\label{massiveV}
\eeq

\begin{figure}[ht]
\includegraphics[scale=.48]{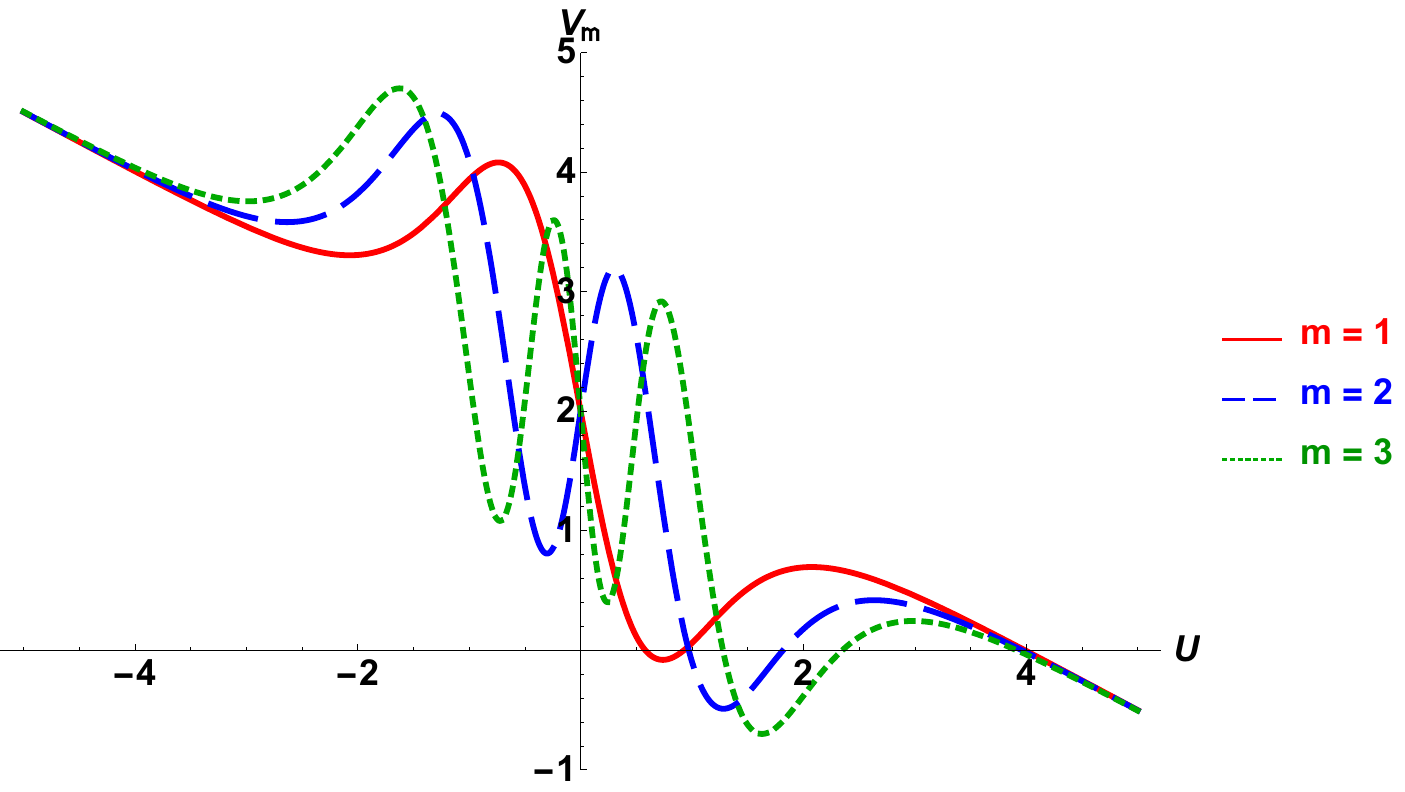}
\vskip-3mm\caption{\textit{\small For non-zero relativistic mass ${\fm}$ the lightlike ``vertical'' trajectory $V_{\fm}(U)$ becomes tilted, as shown for ${\fm}=M=1$.
}
\label{Vmplot1}
}
\end{figure}

Do we get DM also in the massive case ? At first sight, the answer seems to be \emph{negative}~:
the linear-in-$U$ term tilts the vertical coordinate \eqref{massiveV}  as shown in FIG.\ref{Vmplot1}.

However switching  to the longitudinal coordinate
\footnote{Choosing  instead
$
\tilde{Z}_{\fm} = V +\half \frac{\fm^2}{M^2}\,U\,
$
would work for DM however it would not preserve the form of the metric.},
\beq
Z = V +\half U\,,
\label{ZVU}
\eeq
we get
$$
Z = V_0+\frac{1}{2}\left(1-\frac{\fm^2}{M^2}\right)\,U\,,
$$
and thus $U$ is eliminated when
\beq
\fm = M\,.
\label{fmM}
\eeq
Then the trajectory is tipped back to horizontal, leaving us with
\beq
Z(U) \equiv Z_{{\fm}}(U)= V_0 = \const
\label{Zmfix}
\eeq

\begin{figure}[ht]
\includegraphics[scale=.4]{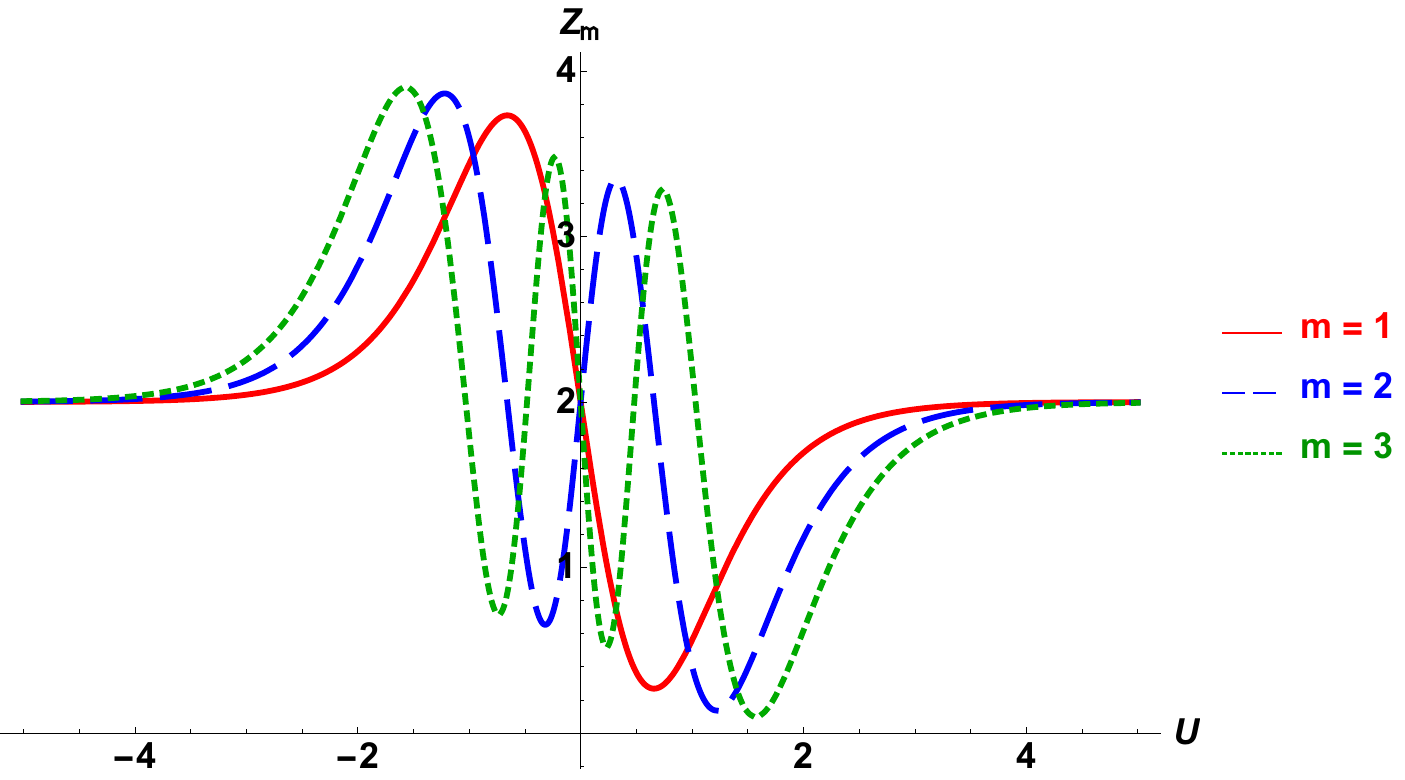}
\vskip-3mm\caption{\textit{\small When the two types of masses are equal, \eqref{fmM}, then swapping the light-cone coordinate $V_{\fm}$ for the  longitudinal one, $Z_{\fm}$ in \eqref{ZVU}, tips the trajectory back which becomes similar to $\hat{V}(U)$ for ${\fm}=0$ in FIG.\ref{PT-XV}.
}
\label{Zmplot1}
}
\end{figure}

In conclusion, we do obtain \emph{DM for all coordinates provided the relativistic and the non-relativistic masses are equal}, \eqref{fmM}. The transverse trajectory $X(U)$ is independent of $V$  and the effect of $\fm$ amounts to merely replacing $\hat{V}$ in FIG.\ref{PT-XV} by
 $Z_{\fm}$ in FIG.\ref{Zmplot1}.

The results of this section can also be presented from a slightly different point of view, see the Appendix.

\section{Carroll-symmetry-related approach }\label{CarrollSec}

In  our  paper \cite{LongMemory} we argued that \emph{no permanent displacement is possible} --- whereas here  we have just found two counter-examples. Now we clarify how does this come about.

First of all, we have shown in the previous sections that the relativistic mass $\fm$ has no effect on the transverse dynamics and merely adds a linear-in-$U$ mass term $-({\fm^2}/{2M^2})U$, \eqref{VcmM}, to the $V$-dynamics \cite{EDAHKZ}. Swapping the light-cone coordinate $V$ for the longitudinal $Z$ as in \eqref{ZVU} the problem then takes the same form as for $V$. Therefore we restrict henceforth our attention to lightlike geodesics, $\fm=0$, with NR mass parameter scaled to $M=1$ for simplicity.
  Then the Carroll-symmetry-based approach  \cite{Carroll4GW,ShortMemory,LongMemory,Leblond} provides us with the following road map:

\begin{itemize}
\item
 We choose an $U_0$ and solve the \SL equation for the  $1\times1$ ``matrix'' $Q$~\footnote{
In $D$-dimensional transverse space $Q$ is a $D\times D$ matrix \cite{Sou73,PolPer}. The initial conditions in \eqref{QSLeq} can be modified, as discussed e.g. in \cite{EZHRev}.}{\,},
\beq
\ddot{Q} + \half{\cA}\,Q=0 \quad
\text{\small with initial conditions}\quad
Q(u_0)=1, \quad \dot{Q}(u_0)=0\,.
\label{QSLeq}
\eeq

\item
Using $Q$ we switch from Brinkmann (B)
to Baldwin-Jeffery-Rosen (BJR) coordinates \cite{BaJeRo,Gibbons75},
\beq
X=Q\, x, \quad U= u,\quad
V = v -\smallover{1}/{4}\,\dot{a}\, x^2\, \where a(u) = Q^2(u)\,.
\label{BJRfromB}
\eeq
$(X,U,V) \to (x,u,v)$ carries the metric \eqref{Bmetric}
 to the BJR form \cite{BaJeRo,Sou73,Gibbons75,SLheart,EZHRev},
\beq
a(u)dx^2 + 2du dv\,.
\label{BJRmetric}
\eeq
\item
The requirement of \emph{being at rest in the Beforzone} i.e.
 before the sandwich wave arrives implies \cite{ShortMemory} that the transverse BJR  trajectory is trivial,
\beq
x(u) = x_0\,, \qquad
v(u)= v_0 \, .
\label{trivtraj}
\eeq
\item
Then the B-trajectory is,
\beq
X(U)= Q(U)x_0\,,\qquad
V(U) = v_0 - \frac{1}{2}({Q}\dot{Q})x_0^2\,.
\label{XUQVtraj}
\eeq

\end{itemize}

\smallskip
This road map allowed us to conclude  that \emph{no permanent displacement is possible}~: we have VM, but no DM \cite{LongMemory}. Anticipating the details to come,
 we remind the reader of that while Brinkmann coordinates are
 defined for all $U$, the BJR coordinates are defined only in coordinate patches $I_{k}= (u_k,u_{k+1})$~\cite{Sou73,Carroll4GW,ShortMemory} \footnote{The singularity of BJR coordinates was instrumental in the long-standing controversy about the physical existence of \GWs \cite{Bondi57}.},
distinguished by
\beq
a(u_k)=0\quad\Leftrightarrow\quad Q(u_k)=0\,,\quad k\geq 1 \;\;\text{integer}\,.
\label{azeros}
\eeq
The BJR coordinates become singular at the end points points. The B $\Leftrightarrow$ BJR correspondence  \eqref{BJRfromB}  works in each $I_k$ separately but should be matched at the contact points.
Crossing such a contact point $u_k$ one has to restart the B $\Leftrightarrow$ BJR transcription -- however
 the simple formulas  \eqref{trivtraj}-\eqref{XUQVtraj}, which are valid for \emph{zero initial velocity}, have to be replaced by a considerably more complex procedure
 \cite{Sou73,Carroll4GW}.

\smallskip
In detail, comparing \eqref{QSLeq} with \eqref{geoX} shows that
 $Q(U)$ satisfies the same equation \eqref{geoX} as $X(U)$ does for initial conditions $X(U_0)=1$ and $\dot{X}(U_0)=0$. The  geodesics are conveniently found by switching to BJR and by using the conserved quantities generated by the
 symmetries of plane gravitational waves which, in addition to  translations, involve also Carroll boosts \cite{Leblond,Sou73, Carroll4GW}. These symmetries leave $u$ fixed  and act on the BJR coordinates as,
\beq
x\to  x + S(u)b+c,
\aand
v \to v - b\,x -\half S(u)\,p^2 + f,
\label{Carrollact}
\eeq
where $c,\,b,\,f$ are real numbers and
\beq
 S(u)=\int_{u_0}^u\!\!\frac{du}{a(u)}\,
 \label{Smatrix}
 \eeq
is the Souriau ``matrix'' \cite{Sou73,Carroll4GW} (a scalar for our $D=1$). \eqref{Carrollact}
 preserves the BJR metric \eqref{BJRmetric} and generates by Noether's theorem conserved linear and Carroll momenta,
\beq
p = a(u)\,\dot{x}(u),
\quad
M=1,
\quad
k=x(u)-S(u)p\,,
\label{CarCons}
\eeq
respectively. Conversely, these conserved quantities determine the BJR trajectories \cite{Carroll4GW},
\beq
 x(u)= k + S(u) p\,,
 \qquad
v(u)=v_0-\half S(u)\,p^2 \,,
\label{genBJRtraj}
\eeq
where  $v_0=v(u_0)$.
\goodbreak

Things are particularly simple when the incoming velocity
 is zero as it is required in the Beforezone. Then
\beq
0 = \dot{X}(U_0) =\dot{Q}(U_0)x_0+Q(U_0)\dot{x}(U_0)=
\dot{x}(U_0)\,,
\label{invel}
\eeq
which by \eqref{CarCons} implies the vanishing of the incoming conserved momentum,
\beq
p=0\,.
\label{pzero}
\eeq
 The Souriau-terms in \eqref{genBJRtraj} are then switched off and the trajectory is  just a fixed point,
  \eqref{trivtraj} with $x_0=k$, and the globally defined B-trajectory is  recovered
 by pulling it back to Brinkmann by \eqref{BJRfromB}\cite{ShortMemory,EZHRev}.
However  when the  momentum does {not} vanish,
$
p\neq0\,,
$
then \eqref{trivtraj} and thus \eqref{XUQVtraj} are not more valid and should be replaced by  \eqref{genBJRtraj} which requires to find $p$ and then to calculate the Souriau matrix.

We emphasise that all our investigations above and \eqref{trivtraj} in particular are valid only where the BJR coordinates are valid, \ie,
 in the interval $I_k$
 between two subsequent zeros of $Q(u)$. Then we are left with the task of gluing together the results obtained in neighboring domains $I_k$.

\kikezd{Illustration: \PT in BJR}

\begin{figure}[ht]
\includegraphics[scale=.42]{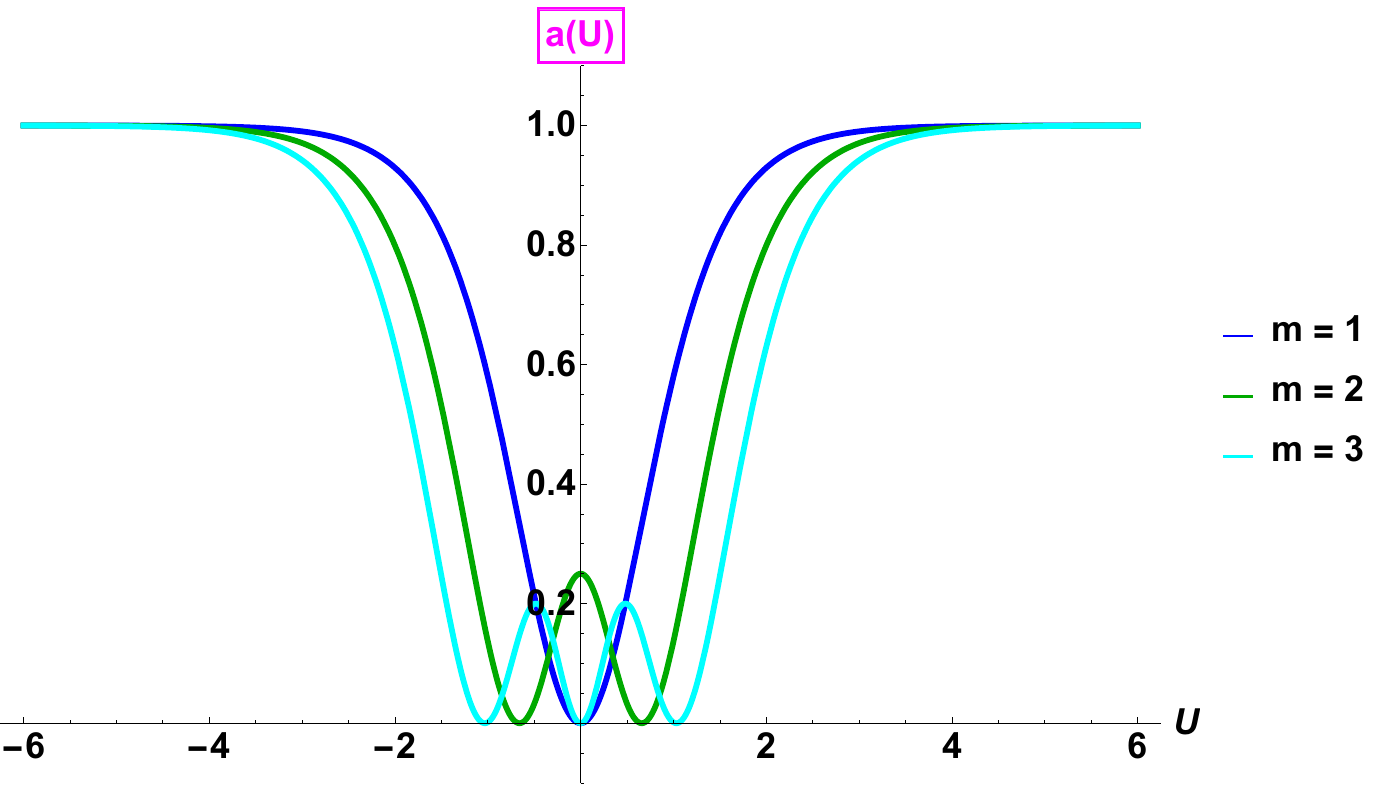}
\vskip-3mm\caption{\textit{\small The BJR profile for the \PT potential for  wave numbers \magenta{${\bf m=1, 2,3}$} }
\label{PT-BJR-int}
}
\end{figure}

More insight is gained by illustrating the procedure by considering
the \PT potential \eqref{PTPot}-\eqref{PTkm} with $M=1$, for which we had found the analytic solutions $X(u)=-P_{m}\left(\tanh u\right)$ in \eqref{PT-XmU}.
The BJR profile, found by following our road map
is plotted in  FIG.\ref{PT-BJR-int}. It has $m$ zeros, and the procedure has to be restarted in each of the intervals $I_{k}$.

\begin{figure}[ht]
\includegraphics[scale=.33]{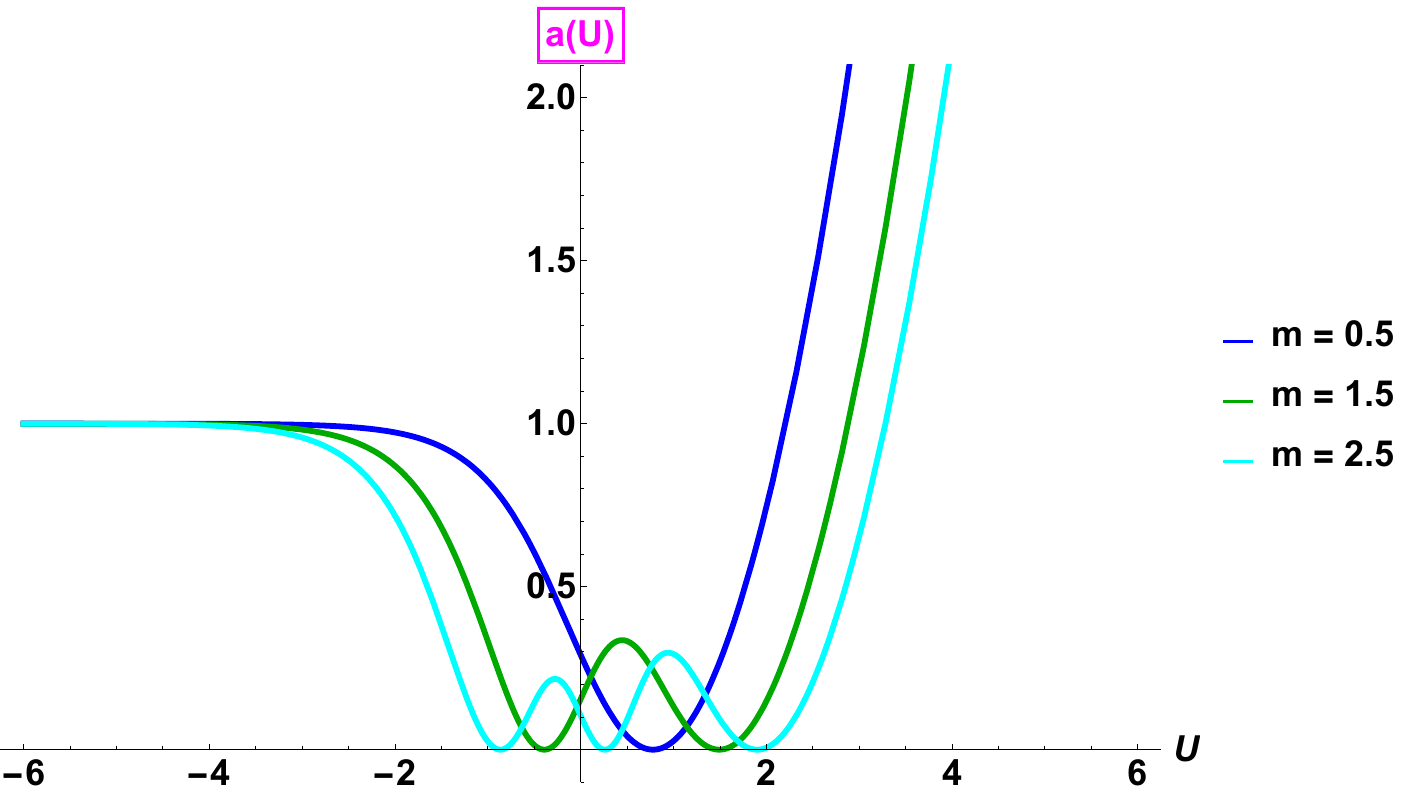}\\
\vskip-2mm\caption{\textit{\small The BJR profile for the \PT
potential for  ${\bf \magenta{m}}$ is not an integer. The force falls off before the velocity is brought down to zero and then the particle flies off with  constant velocity.
\label{PT-BJR-frac}}
}
\end{figure}

\begin{figure}[ht]\hskip-2mm
\includegraphics[scale=.4]{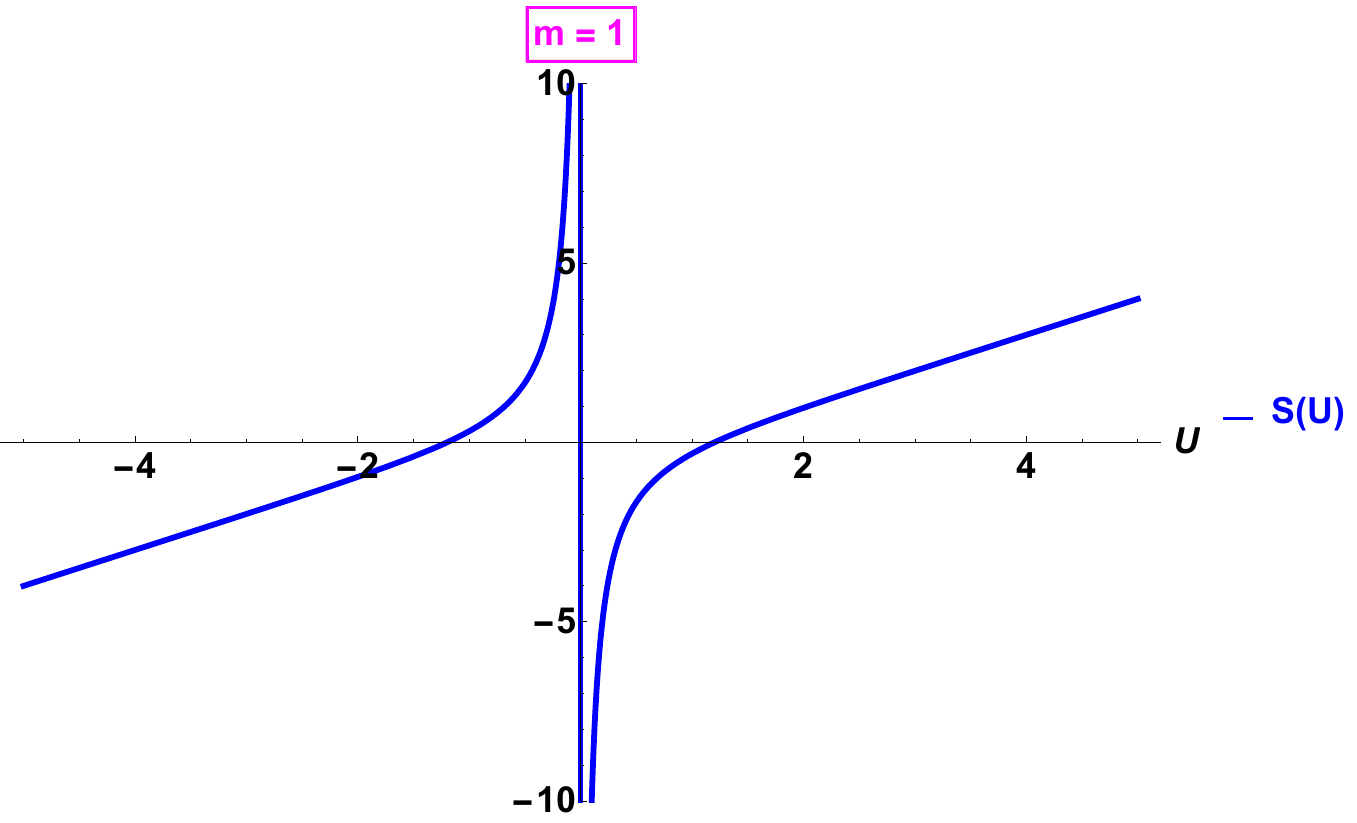}
\vskip-4mm
\caption{\textit{\small The Souriau matrix $S_{m=1}$ in
\eqref{Sm1matrix} is regular and approximately linear outside the wave zone both in $I_{-}$ and $I_{+}$ but diverges at their junction at $u_1=0$.
}}
\label{Sm1fig}
\end{figure}

\begin{figure}[ht]\hskip-2mm
\includegraphics[scale=.355]{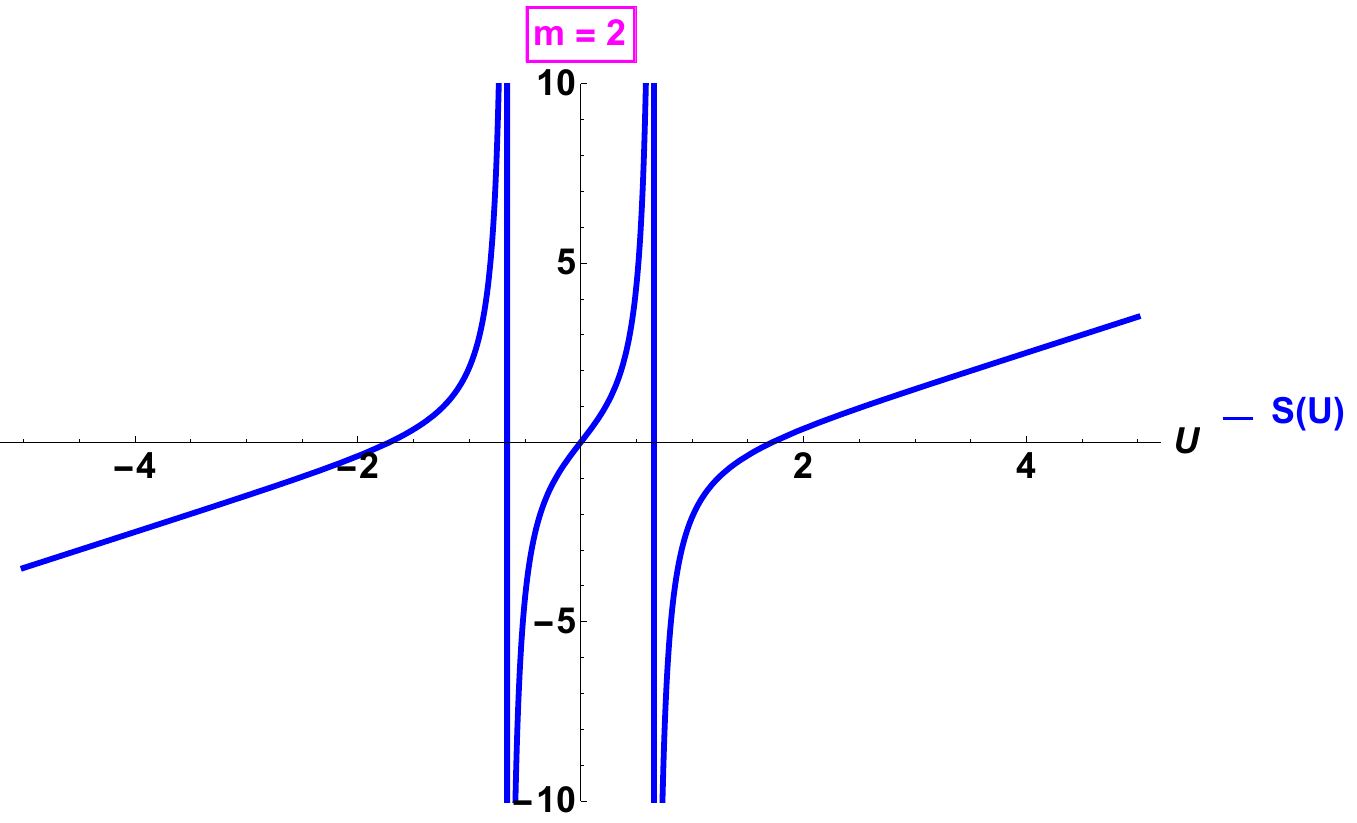}
\includegraphics[scale=.355]{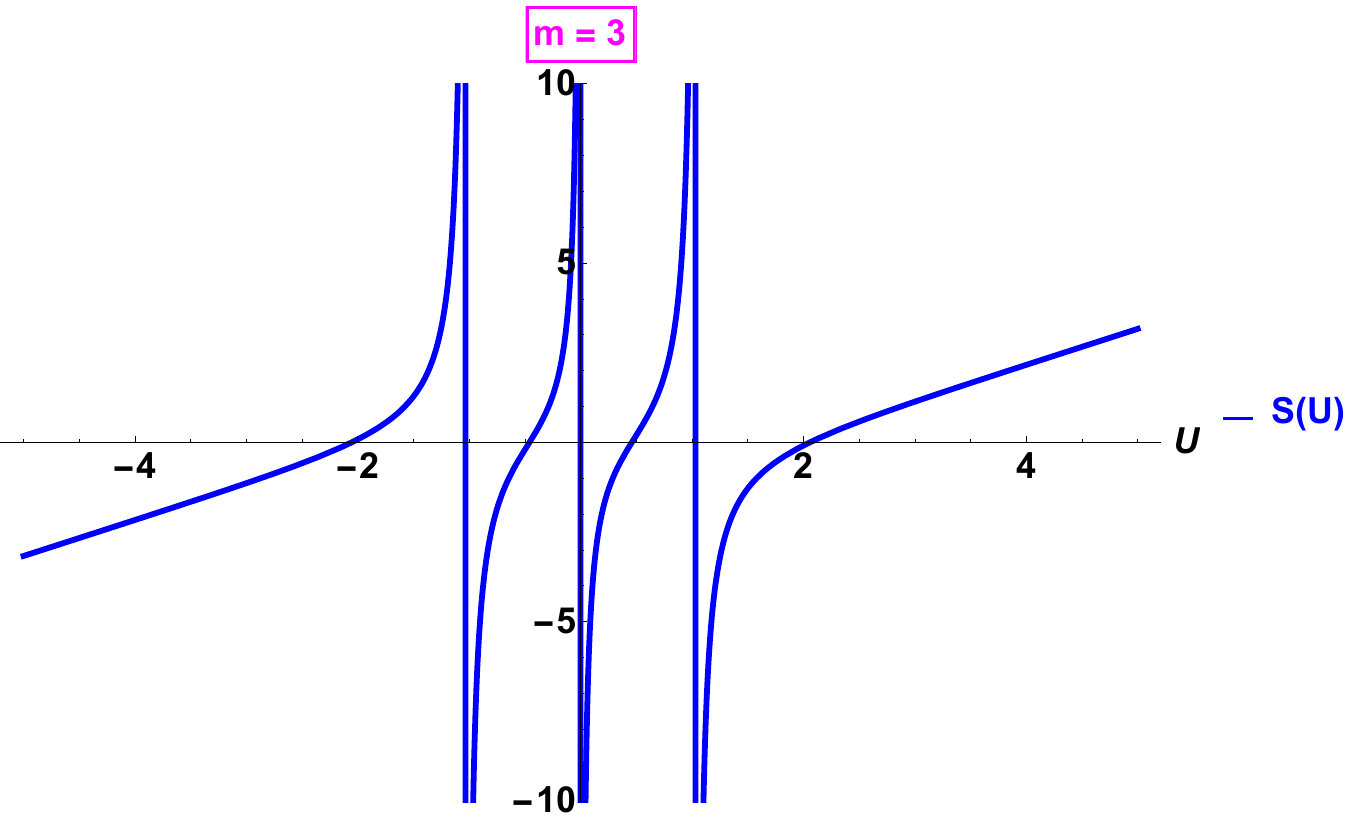}
\vskip-4mm
\caption{\textit{\small The  Souriau matrices
for wave numbers $\magenta{\bf m=2,\, 3}$ are regular in $m+1$ domains separated by the zeros  of the B-trajectory,  where they diverge.
}}
\label{Sm2m3fig}
\end{figure}

We start with a  trajectory $X(U)$ viewed as $Q$ ``matrix''.
For an integer wave number $m$ the metric \eqref{BJRmetric} becomes, outside the wave zone,  that of Minkowski, FIG.\ref{PT-BJR-int} (whereas it diverges   for $u\to \infty$ when $m$ is fractional FIG.\ref{PT-BJR-frac}).
%
The Souriau matrix \eqref{Smatrix}, %
\begin{equation}
S(u) \equiv S_{m}(u) =
\int_{u_0}^u\!\!\frac{du}{\big[P_{m}(\tanh u)\big]^{2}}\,,
\label{PTSmatrix}
\end{equation}%
is well-defined between two subsequent zeros of the denominator  which are indeed those of the Legendre polynomial in sect.\ref{PTSec}.
For the $m=1$, for example, the Brinkmann trajectory $Q(U)=X(U)=-\tanh U$ shown in FIG.\ref{PT-m1} yields,
\begin{equation}
S(u) \equiv
S_{m=1}(u)=u-\coth u\,,
\label{Sm1matrix}
\end{equation}
depicted in FIG.\ref{Sm1fig}. For $m=2$  we have instead,
\beq
S_{m=2}=u+\frac{3\sinh \left(2u\right)}{4-2\cosh \left(2u\right)}\,.
\label{Sm2matrix}
\eeq
Both in the Before and in the Afterzone, the Souriau matrix is approximately linear, as shown in FIGs. \ref{Sm1fig} and \ref{Sm2m3fig}.
 Focusing our attention henceforth at  $m=1$, we note that the Souriau matrix $S_{m=1}(u)$ is regular in both of the two domains
\beq
I_{-} = (-\infty,0)
\aand
I_{+} = (0,\infty)\,
\label{BJRdomains}
\eeq
but diverges at $u=0$ --- the point where the  Brinkmann trajectory vanishes.

Eqn. \eqref{genBJRtraj} is valid \emph{separately} in both coordinate patches $I_{\pm}$,
  \besub
\begin{align}
x^{\pm}(u)= &\;k^{\pm} \;+ \; p^{\pm}\Big(u-\coth u\Big)
\,,
\qquad
&v^{\pm}(u)= \;v_0^{\pm} \;- \; \frac{(p^{\pm})^{2}}{2}\Big(u
-\coth u\Big),
\label{BJRm1xv}
\\[8pt]
\dot{x}^{\pm}(u) = &\; p^{\pm}\left(1+\frac{1}{\sinh^2 u}\right)\,,
\qquad\;
&\dot{v}^{\pm}(u) = -\frac{(p^{\pm})^2}{2}\left(1+\frac{1}{\sinh^2 u}\right)\,,
\label{PT-BJR-vels}
\end{align}
\label{-xv-vels}
\esub
where $k^{\pm}$ and $p^{\pm}$ are arbitrarily chosen constants.
Both  BJR trajectories are regular in their respective domains
\eqref{BJRdomains}, but diverge at $u=0$.
However, pulling the BJR trajectory back to Brinkmann by \eqref{BJRfromB} \emph{removes the singularity}~:
\beq
X(U)= k^{\pm}\tanh U + p^{\pm}\big(U\tanh U - 1\big)\,
\label{pullbacktoB}
\eeq
is regular. Then
\benu
\item
The two BJR branches match at $u=0$ when $p^{+}=p^{-}$~.
\item
From
\beq
\dot{X}(U)=\frac{k^{\pm}}{\cosh^2 U}
+p^{\pm}\left(\tanh U +\frac{U}{\cosh^2 U}\right)
\label{pmvelocity}
\eeq
we deduce that no motion in the Beforezone,
$\dot{X}(-\infty)=0$, requires $p^{-}=0$ and thus
$
p^{+}=p^{-}= p = 0
$
 as in \eqref{pzero}. The Souriau term is thus switched off.
\item
By \eqref{pmvelocity}
 the velocities match when $k^{+}=k^{-}=k$.
\eenu

In conclusion, for $k=-1$ we recover the solution
\eqref{PT-XmU} with $m=1$ \ie,
$
X_1(U)=-\tanh U\,,
$
shown in FIG.\ref{PT-m1}.
At last, \eqref{XUQVtraj} is
\beq
V(U) =V_0- \half\frac{\sinh U}{\cosh^3 U}X_0^2 \;\to\; V_0
\label{VU}
\eeq
consistently with \eqref{m1m2V}.

\begin{figure}[ht]
\includegraphics[scale=.4]{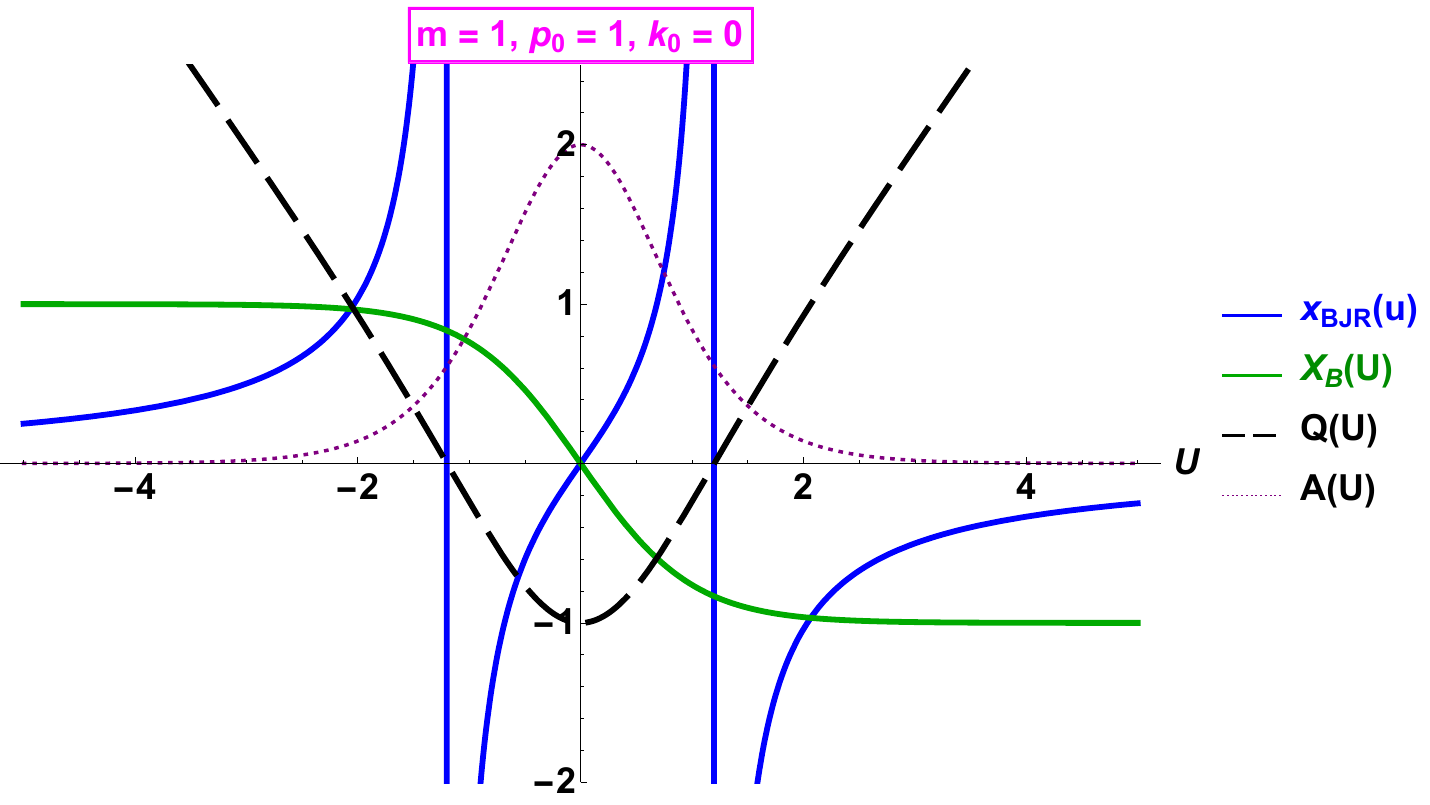}\,
\vskip-3mm\caption{\textit{\small Trajectories for  $\tilde{Q}$ in \eqref{tildeQ} with wave number \magenta{${\bf m=1}$} in \dgreen{Brinkmann} and in \blue{BJR} coordinates.}
\label{B-BJR-X2}
}
\end{figure}

We mention that the initial conditions in \eqref{QSLeq} could actually be modified:
 the important condition for $DM$ is  \eqref{DMcond}.
We illustrate this point by considering
\beq
\tilde{Q}(u) = u \tanh u -1
\label{tildeQ}
\eeq
which  is yet another solution of the \SL equation \eqref{QSLeq}. It  is singular where $u\tanh u=1$ and diverges at $\pm\infty$.
Then our road map above yields the Souriau matrix
\beq
\tilde{S}(u) = \dfrac{\tanh u}{1- u\tanh u}
\eeq
and from \eqref{genBJRtraj} we deduce that,
\beq
\dot{x}(u) = \dfrac{1}{[u\tanh u - 1]^2} \,p
\Rarrow
\dot{x}(u = \pm \infty) = 0\,.
\eeq
Thus the  condition \eqref{DMcond} is satisfied, and we do get DM, as shown in FIG.\ref{B-BJR-X2}.

\section{Conclusion}\label{Concl}

Particles at rest before the arrival of a sandwich  \GW exhibit, generically, the velocity effect (VM)~: the particles fly apart with diverging constant but non-zero velocity \cite{Ehlers,Sou73, GriPol,AiBalasin,ShortMemory,LongMemory, PolPer,SLheart,EZHRev}.
Zel'dovich and Polnarev suggested instead \cite{ZelPol}  that flyby would generate (approximately) pure displacement ({DM}).

Our paper answers a question of the (unknown) referee of our previous paper \cite{EZHRev} concerning the relation of VM and DM. In detail, we argue that for a judicious choice of the wave parameters, namely when the  Wavezone contains an \emph{integer number of  half-waves} then we \emph{do get pure displacement}.
 We  illustrated our statement both by a numerical (sec.\ref{GaussSec}) and analytical (sec.\ref{PTSec}) examples.

Our results can be understood from yet another point of view \cite{TDcorr}.
In the linear approximation,
\beq
X(U)= X(0) + U \dot{X}_0\,
\aand\,
\dot{X}(U) = \dot{X}_0\,,
\label{linear}
\eeq
where $\dot{X}_0$ is the initial velocity. After the wave had passed, the displacement and the velocity depend on three moments,
\beq
M_0=\int\!\cA(U) dU \,, \quad
M_1= \int\! U \cA(U)dU\,,  \quad
M_2= \int\!U^2 \cA(U)dU \,.
\label{moments}
\eeq
Then the motion is
\besub
\begin{align}
&\dot{X}_{sol}(U) = M_0 X_0 + \dot{X}_0 + M_1 \dot{X}_0\,,
\label{solvel}
\\
&X_{sol}(U) = X_0 + U \dot{X}_{sol}(U) - M_1 X_0 - M_2 \dot{X}_0\,,
\label{solX}
\end{align}
\label{solvelX}
\esub
where ${X}_0={X}(U_a)$ and $\dot{X}_0=\dot{X}(U_a)$
are the initial position and velocity.
In the Afterzone $U > U_a$ where $\cA(U)\equiv0$ all three moments vanish, $M_0 = M_1 = M_2 = 0$. Thus the motion is along straight lines with constant velocity \cite{PolPer}, which  vanishes if the incoming particle had zero velocity,
$ \dot{X}_0 = 0$. But then it stops for good, \beq
X_{sol}(U) = X_0=X(U_a) = \const\, \for U> U_a\,.
\label{XsolX0}
\eeq
  The difficulty is to find out for which values of the parameters does this happen. Our answer given to this question is~: one should have an integer number of half-waves.
\goodbreak

Our new results complete those previous ones \cite{ShortMemory,LongMemory,PolPer,SLheart,EZHRev}
which are valid in a domain where the BJR coordinates are regular. The question is studied in detail  and illustrated
for the \PT profile in sec.\ref{CarrollSec}.
\goodbreak

Generalization to  4 dimensions
with physical applications will be studied further in \cite{DMvsVM}, confirming that reducing {VM} to {DM} is indeed possible  also for more general plane gravitation waves which include those generated by flyby, gravitational collapse, etc as proposed in \cite{GibbHawk71}.

\begin{figure}[ht]
\includegraphics[scale=.22]{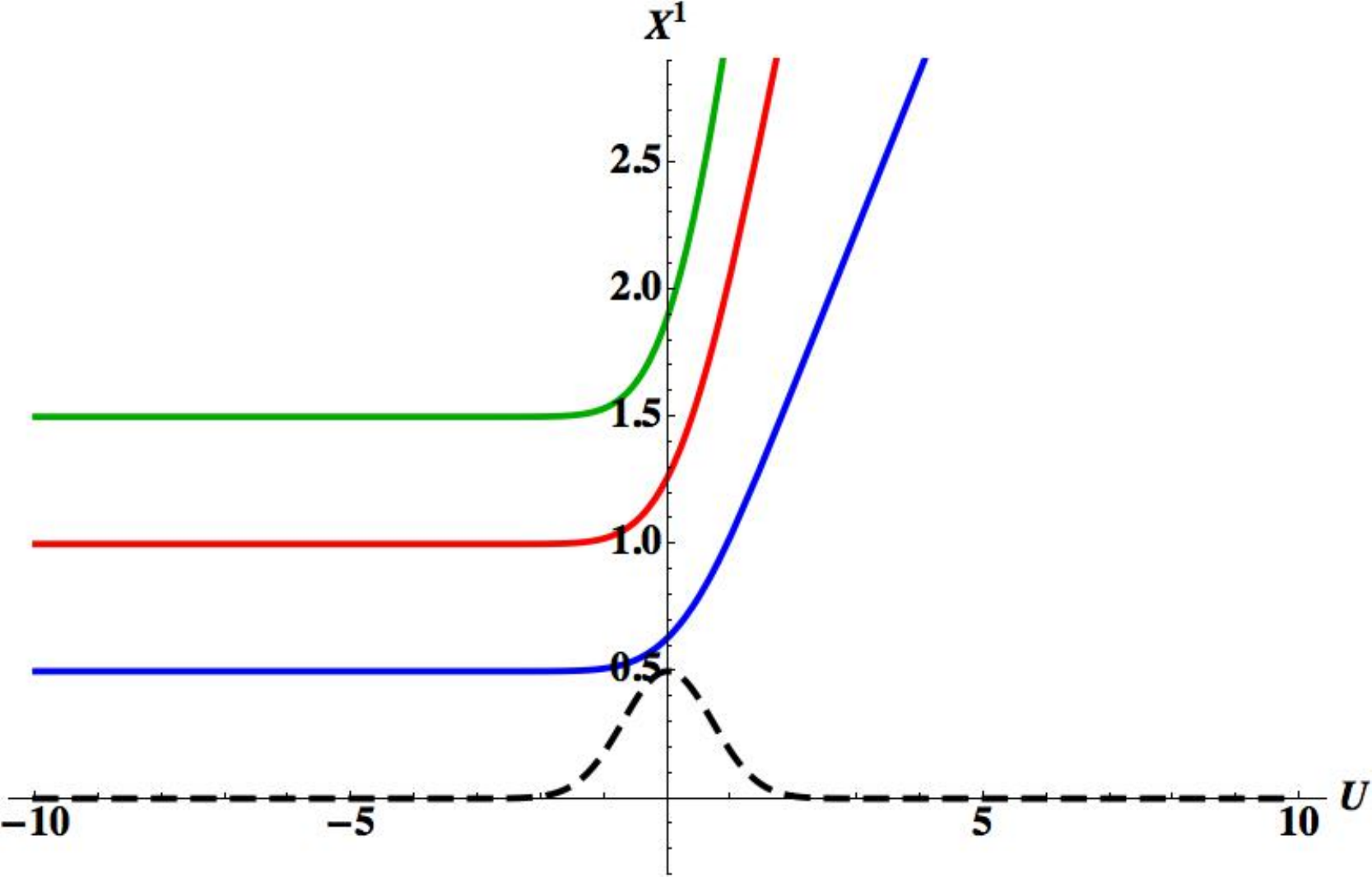}\qquad\quad
\includegraphics[scale=.23]{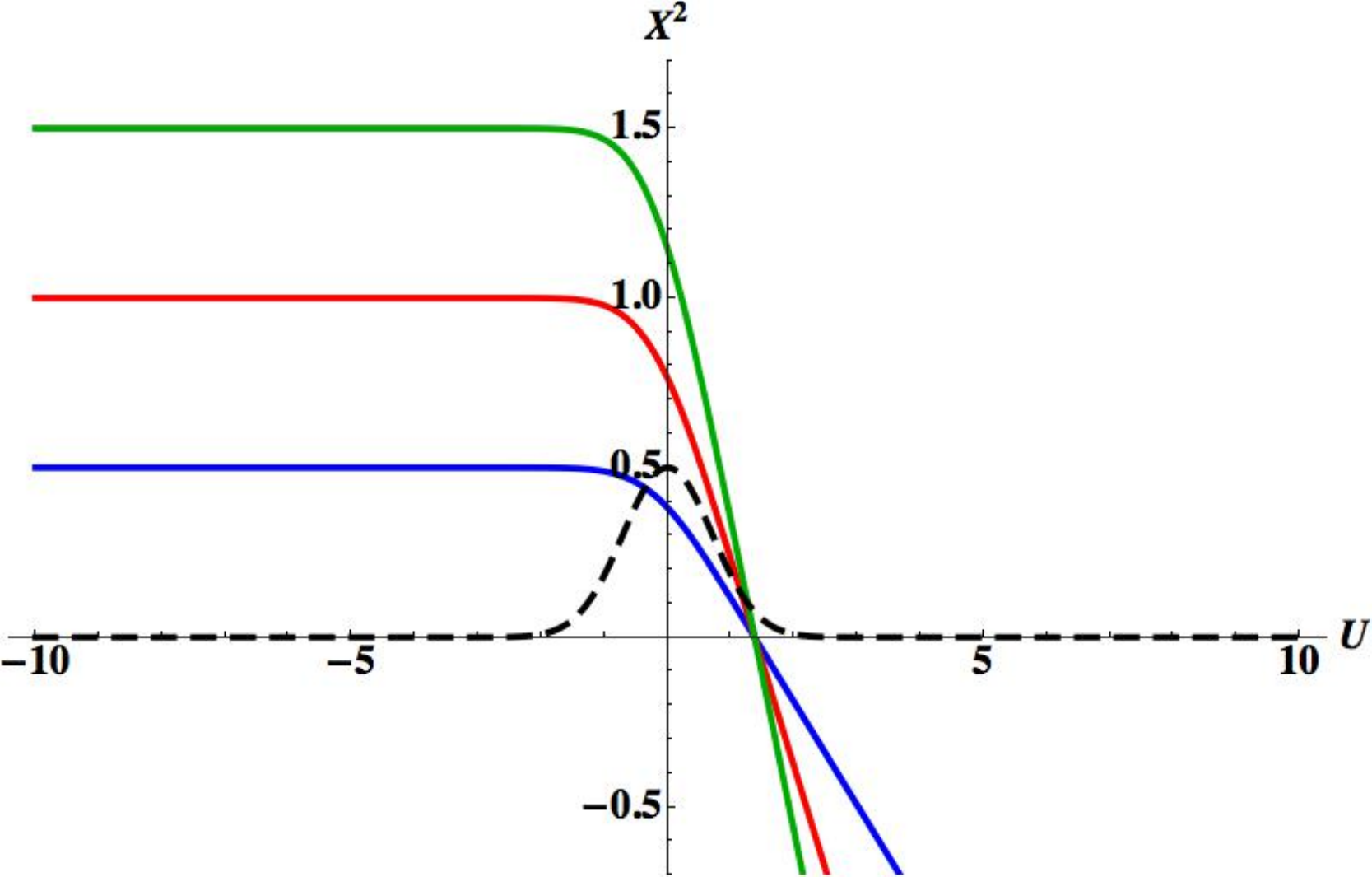}
\vskip-3mm
\caption{\textit{\small The geodesics  for the Gaussian burst \eqref{A0G} in  2 transverse dimensions for various (\blue{blue}/\red{red}/\dgreen{green})~ initial positions in the Beforezone.
The potential is attractive in the $X^2$ but repulsive in the $X^1$ sector, implying diverging trajectories in the latter.
}
\label{2DGausstraj}
}
\end{figure}

Both numerical and analytical evidence show, for example, a particular behavior for vacuum \GWs with a Brinkmann metric
with 2 transverse direction
\beq
g_{\mu\nu}dX^\mu dX^\nu=
\delta_{ij} dX^i dX^j + 2 dU dV +
\half{\cA}(U)\Big((X^1)^2-(X^2)^2\Big)dU^2\,
\label{2Bmetric}
\eeq
\!where $(X^i), \, i=1,2$ are transverse and $U,\, V$ light-cone coordinates. The relative minus here
is mandatory for a vacuum \GW\!, therefore one of the components necessarily diverges as shown
by FIG.\ref{2DGausstraj} [FIG.3 of \cite{LongMemory}] for the Gaussian profile  \eqref{A0G} with $k=1$.
When looking for {DM}, the diverging coordinate should be discarded by putting it to identically zero, allowing for a ``half {DM}'' after fine-tunig. 

\begin{figure}[h]
\includegraphics[scale=.5]{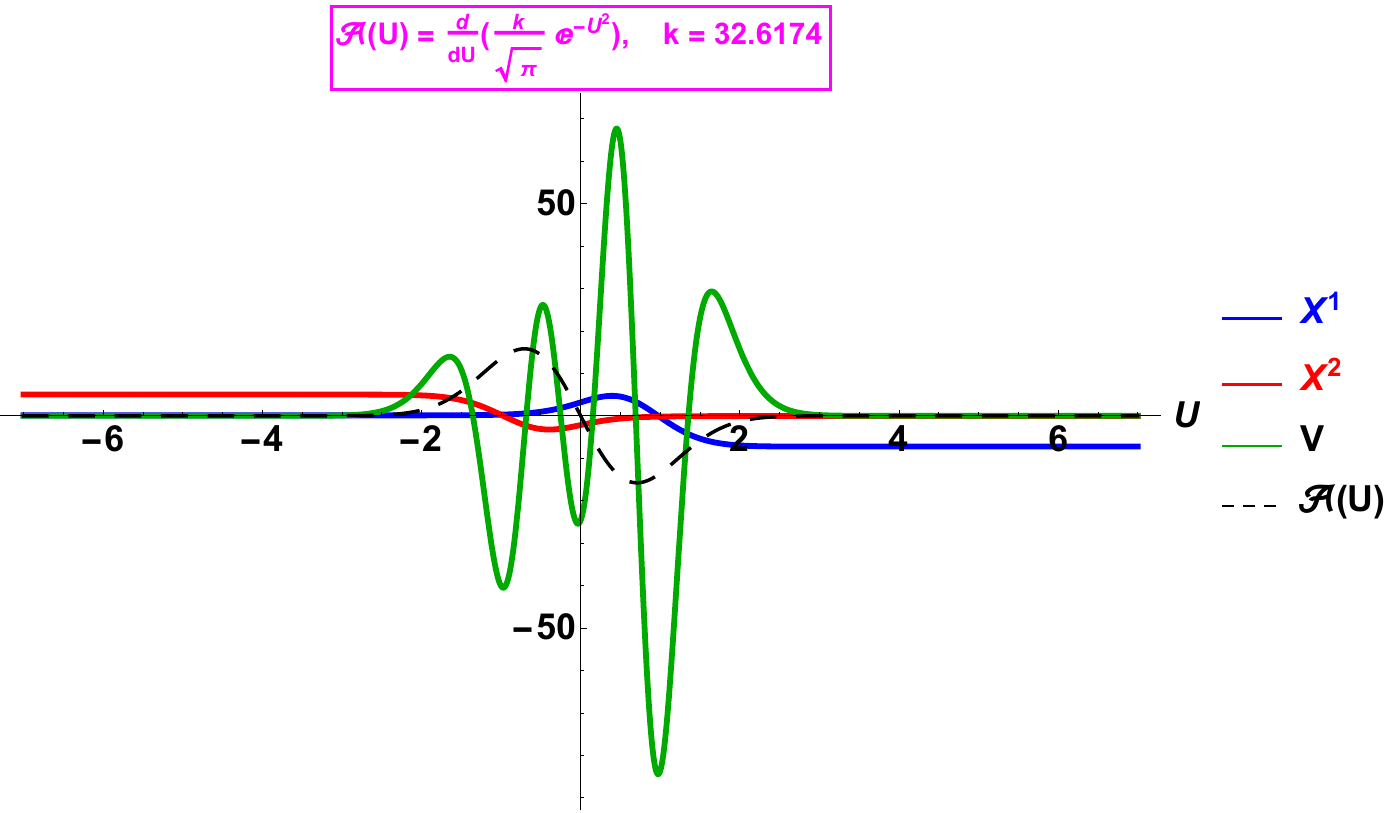}
\\
\vskip-4mm
\caption{\textit For 
flyby  with \magenta{$m=1$} half-wave we have full DM for  all three components. 
\label{d1-Gauss-xyv}
}
\end{figure}

We stress however that this ``halfening'' is unrelated to the question if our wave is a vacuum \GW or not, but depends rather on the profile. Following \cite{GibbHawk71}, flyby should be described, for example, by a vacuum \GW\! whose profile is the first derivative of the Gaussian. And it has full DM for both coordinates, as shown in FIG.\ref{d1-Gauss-xyv}.

This behavior corresponds  indeed to a general pattern: {even-order} derivatives of the Gaussian exhibit
 half {DM} and {odd-order} derivatives exhibit full
{DM} in both transverse directions  \cite{DMvsVM}.

 Related results were discussed in \cite{Chakraborty:2022qvv,Mitman,BiPol} and more recently in \cite{Jibril,HarteOancea}.

\vskip-6mm
\begin{acknowledgments}\vskip-4mm
 We are grateful to the anonymous referee of our previous paper \cite{EZHRev} for directing our interest at the VM vs DM problem. We are indebted to G.~Gibbons and to J. Balog for their insights and advices.
The Appendix was contributed by J. Balog. Correspondence
 is acknowledged to T. Damour and G. Junker.  We profited from
discussions with P.~C. Aichelburg,  L. Di\'osi and M. Elbistan, Z. Silagadze and F. Ziegler. PAH thanks the Schr\"odinger Institute (Vienna) for hospitality during the Workshop {\sl Carrollian Physics and Holography}${}_{-}$CDFG${}_{-}$2024 in April 2024. PMZ was partially supported by the National Natural Science Foundation of China (Grant No. 12375084).
\end{acknowledgments}
\goodbreak


\goodbreak

\newpage

\bigskip
\appendix
\section{\bf Geodesic motion of a massive particle
}\label{Appendix}

The relativistic Lagrangian for geodesic motion in $D=1$ transverse direction is, in Brinkmann coordinates,
\begin{equation}
{\cal L}_{geo} =
(\dot X)^2+2\dot U \dot V -\half \cA(U)\,X^2\,\dot U^2\,,
\end{equation}
where the dot denotes derivation w.r.t. an affine parameter
$\lambda$. The Euler-Lagrange equations are,
\begin{equation}
\begin{split}
\ddot X&=-\half\cA\,X\,\dot U^2,\\
\ddot U&=0,\\
\ddot V&=\frac{1}{4}\frac{d{\cA}}{dU}\, X^2\,\dot U^2+\half\cA \dot{(X^2)}\,\dot U\,.
\end{split}
\end{equation}
 The $U$ equation is integrated at once,
yielding the non-relativistic mass familiar in the Bargmann framework \cite{DBKP,DGH91}, $\dot U = M = \const$

The Euler-Lagrange equations then imply that the Lagrangian is conserved along the geodesic:
$
\frac{\;d}{d\lambda}{\cal L}_{geo}=0
$
providing us with a constant of the motion,
\beq
{\cal L}_{geo}=-\frac{\fm^2}{M^2}\,,
\label{Lgeoconst}
\eeq
where $\fm^2$ is the Jacobi invariant \eqref{Jacobiinv}.
Switching to longitudinal and relativistic time coordinates,
\begin{equation}
z=V+\half U \aand t=V-\half U\,,
\end{equation}
respectively, three essentially different cases can be distinguished,
\beq
\left\{\barraynb{lll}
\fm^2 > 0 &\text{timelike geodesic}  &\text{for massive particle}
\\
\fm^2=0 &\text{lightlike geodesic} &\text{for massless particle}
\\
\fm^2 < 0 &\text{spacelike geodesic} &\text{for tachyonic particle}.
\earraynb \right.
\eeq
\noindent
Dropping tachyons we consider henceforth  $\fm^2\geq0$
and scale $M$ to $1$~.

\smallskip
Replacing the affine parameter by $U$ and denoting  ${\rm d}/{\rm d}U$ by prime, the transversal  Sturm-Liouville  equation, \eqref{geoX}
\begin{equation}
X^{\prime\prime}=-\half\cA\,X\,,
\label{SLeq}
\end{equation}
 is obtained; integrating twice  the $V$ equation \eqref{geoV}    yields,
\begin{equation}
z(U)=V_0 +\half\left(1-\left(\frac{\fm}{M}\right)^2\right)U
-\half X{\,}X^{\prime}\,.
\label{zUX}
\end{equation}
\goodbreak

For a particle initially  at rest in the
Before zone we must have
$z=\const$
Then using that $X^{\prime}=0$ in the Before zone,
allows us to conclude that DM in the Afterzone requires
 that the two types of masses, $\fm$ and $M$ be equal,
\beq
\left(\frac{\fm}{M}\right)^2=1\,
\label{mMXX'}
\eeq
Then \eqref{zUX} reproduces \eqref{Zmfix}.

We underline that our proof does not apply in  the lightlike case $\fm=0$ because the condition \eqref{mMXX'} can not be  satisfied: photons can not be in rest.
Requiring no motion for a massless particle would be unphysical anyway, as said before.

These considerations are completely general in  the conclusion does not depend on the details of the $\cA(U)$ profile and are valid in any transverse dimension including the physically relevant $D=2$.

\end{document}